\documentclass[amsfonts]{article}
\usepackage{graphicx}

\usepackage{subcaption}

\newcommand{\be}{\begin{equation}}
\newcommand{\ee}{\end{equation}}
\newcommand{\bea}{\begin{array}}
\newcommand{\ea}{\end{array}}
\newcommand{\beqa}{\begin{eqnarray}}
\newcommand{\eeqa}{\end{eqnarray}}

\newcommand{\eean}{\end{eqnarray*}}

\def\up#1{\leavevmode \raise.16ex\hbox{#1}}

\newcommand{\gapproxeq}{\lower
 .7ex\hbox{$\;\stackrel{\textstyle >}{\sim}\;$}}
\newcommand{\lapproxeq}{\lower .7ex\hbox{$\;\stackrel
{\textstyle <}{\sim}\;$}}

\newcounter{appendice}

\def\thebibliography#1{{\bf REFERENCES\markboth
 {REFERENCES}{REFERENCES}}\list
 {[\arabic{enumi}]}{\settowidth\labelwidth{[#1]}\leftmargin\labelwidth
 \advance\leftmargin\labelsep
 \usecounter{enumi}}
 \def\newblock{\hskip .11em plus .33em minus -.07em}
 \sloppy
 \sfcode`\.=1000\relax}

\begin{document}
\centerline{ \LARGE  Spinning $\sigma$-model solitons in $2+1$ Anti-de Sitter space }

\vskip 2cm
\centerline{B. Harms\footnote{bharms@ua.edu} and A. Stern\footnote{astern@ua.edu} }
\vskip 1cm
\begin{center}
{ Department of Physics, University of Alabama,\\ Tuscaloosa,
Alabama 35487, USA\\}
\end{center}
\vskip 2cm
\vspace*{5mm}
\normalsize
\centerline{\bf ABSTRACT}
We obtain numerical solutions for rotating topological  solitons of the nonlinear $\sigma$-model in three-dimensional Anti-de Sitter space. Two types of solutions, $i)$ and $ii)$, are found. The $\sigma$-model fields are everywhere well defined for both types of solutions, but they differ in their space-time domains.  Any time slice of the space-time  for  the type $i)$ solution has a causal singularity, despite the fact that all scalars constructed the curvature tensor are bounded functions.  No evidence of a horizon is seen for any of the solutions, and therefore   the type $i)$ solutions have naked singularities. 
 On the other hand,  the space-time domain, along with the fields, for the  type $ii)$ solutions are singularity free.  Multiple families of solutions exhibiting bifurcation phenomena are found for this case. 

\bigskip
\bigskip

\newpage

\section{Introduction}

 Asymptotically $AdS$  solutions are of  current interest due to their application to holography and their possible indication of phase transitions in the boundary field theory.\cite{Witten:1998zw} Examples of such solutions are $AdS$ black holes,\cite{Banados:1992gq} $AdS$ solitons,\cite{Horowitz:1998ha} and their hairy extensions.\cite{Henneaux:2002wm},\cite{Banados:2005hm},\cite{Brihaye:2013tra},\cite{Anabalon:2016izw}  Here we show the existence of asymptotically $AdS^3 $ $\sigma$-model solitons.  Their stability requires    the fields to be rotating.  Nonrotating  asymptotically $AdS^3 $ $\sigma$-model solitons were previously shown  not to exist.\cite{Bizon:2004wa}   This also is evident  from a simple scaling argument. While the $\sigma$-model Lagrangian is scale invariant in two spatial dimensions for static field configurations, this is no longer the case in a background anti-de Sitter space.  Rather, there is a contribution which scales like $r^2$, leading to an attractive force in addition to the gravitational attraction.   The absence of any stabilizing forces, is thus consistent with the nonexistence of static solutions.  The above arguments do not apply for rotating field configurations.  We show that, as a result, there exist rotating  topological solitons which approach $AdS^3$ in the large distance limit.  Asymptotically flat self-gravitating  solitons  in the  $2+1 $ dimensional nonlinear $\sigma$-models have been known to exist for a long time.\cite{Clement:1976hh}.
Analogous self-gravitating solutions, or skyrmions,  in $3+1 $ dimensions are well known.\cite{Heusler:1991xx} Spinning solutions have also been considered.\cite{Ioannidou:2006nn}
 Solutions with large winding number (corresponding to baryon number) have been proposed to model dense  stars.\cite{Glendenning:1988qy},\cite{Piette:2007wd},\cite{Nelmes:2011zz}
Singularities and horizons  can arise  for the  latter solutions in space-times with various cosmological constants.  Such solutions are hairy black holes, and they have been extensively studied.\cite{Heusler:1991xx},\cite{Luckock:1986tr},\cite{Bizon:1992gb},\cite{Kleihaus:1995vq},\cite{Tamaki:2001wca},\cite{Sawado:2004yq},\cite{Brihaye:2005an},\cite{Nielsen:2006gb},\cite{Duan:2007df},\cite{Doneva:2011gx},
\cite{Gibbons:2010cr},\cite{Canfora:2013osa},\cite{Dvali:2016mur},
\cite{Adam:2016vzf},\cite{Gudnason:2016kuu}
 It is of interest to know if  the  $2+1 $ dimensional nonlinear $\sigma-$model also  admits solutions with horizons at finite distances.

Here we examine the standard nonlinear $\sigma-$model coupled to gravity with a negative cosmological constant.  Our ansatz for $\sigma-$model fields in  $2+1 $ dimensions is suitable for the construction of solitons with arbitrary winding number.  Using numerical methods
we obtain two types of rotating soliton solutions with integer winding number.  They are due to the existence of two types of space-time metrics near the origin.  From either space-time metric one gets that  all scalars constructed from the curvature tensor are bounded at the origin.  Nevertheless, the origin is a casual singularity for one case,  which we denote by  $i)$ and not the other,  which we denote by  $ii)$.  The singularity  for $i)$ closely resembles that of a BTZ black hole. Here it is a  naked singularity  because the solutions have no horizons.\footnote{Solutions with  singular metric tensors are of  physical relevance in $2+1$ gravity.  Examples of singularities are the helical and conical singularities appearing in the metric tensor of point masses with, and without spin, respectively. 
 As with our type i) solutions, all scalar quantities constructed from the associated curvature tensor are  bounded everywhere. 
Furthermore, since our type i), as well as the type ii) solutions, are asymptotically AdS, they  are of interest for the AdS/CFT correspondence.  }  There are no space-time singularities (or horizons) for solutions $ii)$, and therefore they are topological solitons. The $\sigma-$model fields are everywhere well defined  for solitons  $i)$ even though the domain has a singularity, i.e., the fields have a well-defined limit at the casual singularity.  Thus the solitons $i)$  are also restricted to distinct  topological sectors. From their asymptotic form at spatial infinity, the solutions can be labeled by the same parameters, namely  mass and angular momentum, as those of a BTZ black hole, in addition to parameters associated with the matter content.  An alternative  mass and angular momentum can be assigned to the solitons using collective coordinate techniques.  Collective coordinate quantization leads to the usual spectrum for a rigid rotor in two spatial dimensions.

  We denote the nonlinear $\sigma-$model fields by $\Phi_a$, $a=1,2,3$, constrained on $S^2$, $\Phi_a\Phi_a=1$.
The action for $\Phi_a$ coupled to $2+1$ gravity
is
\be S=\int d^3 x\sqrt{-g}\,\Bigl(\frac 1{16\pi G}(  R-2\Lambda)-\frac 12 \partial_\mu\Phi_a \partial^\mu\Phi_a+ \lambda (\Phi_a\Phi_a-1)\Bigr)+ S_{GHY}-S_{AdS} \;, \label{srtngactn}\ee
where $G$ is the three-dimensional version Newton's constant (here in dimensionless units), $\Lambda$ is the cosmological constant
and   $\lambda$ is a Lagrange multiplier.  $S_{GHY}$ is the 
Gibbons-Hawking-York term\cite{York:1972sj} on the boundary at  spatial infinity $r\rightarrow\infty $ 
\be S_{GHY}=\frac 1{8\pi G}\int_{r\rightarrow\infty} d^2 x\,\sqrt{-h}\,K\; .\ee
 $h$ is the determinant of the induced metric on the boundary,  and $K$ is the trace of the extrinsic curvature, $K=-\frac 1{\sqrt{-g}}\partial_\mu(\sqrt{-g}\,\hat n^\mu)$, where $\hat n_\mu$ is the unit vector normal to the boundary.  $S_{AdS} $ is the infinite $AdS$ vacuum action, which we subtract off in order  for the gravity contribution to the action to be finite. $\Phi_a\rightarrow$ constant in order for the matter contribution to the action to be finite.  Therefore  just as in flat space the  domain for the nonlinear-sigma model on any time-slice is $S^2$, and topologically distinct field configurations result.  We demand that  $\Phi_a$ has a unique limit everywhere  on $S^2$, including at the point associated with the origin, which  may or may not be a causal singularity.  We label the topological sectors by the winding number
\be n=-\frac 1{8\pi}\int_{x^0={\rm constant}} d^2x\, \epsilon_{abc}\epsilon_{ij}  \Phi_a\partial_i\Phi_b\partial_j\Phi_c\;,\ee
where the integral is on any time-slice and $n$ is  normalized to be an integer. $\epsilon_{abc}$ and $\epsilon_{ij} $ denote totally antisymmetric tensors, and $i,j,..=1,2$ are spatial indices.

In section two we write down the ansatz for the metric tensor and $\Phi_a$ and give asymptotic solutions near spatial infinity and the origin.
Some numerical solutions are presented in section three. Collective coordinate quantization is shown in section four.  The question of the existence of black hole solutions with nonlinear$\sigma$-model hair is examined in section five, while some brief concluding remarks are given in section six.

\section{Asymptotic solutions}
We parametrize the two-dimensional space by polar coordinates $(r,\phi)$, and the time by $t$. 
Our  ansatz for the metric tensor is expressed in terms of three radial functions $A$, $B$ and $\Omega$,
\be ds^2=-A(r) dt^2 +\frac{B(r)}{A(r)}{dr^2}+ r^2\Bigl(d\phi+\Omega(r)dt\Bigr)^2\label{rotatngmtrc}\;,\ee
while the $\sigma$-model fields $\Phi_a$ are written in terms of one radial function $\chi$ and a fixed angular velocity $\omega$,
\be \pmatrix   {\Phi_1\cr\Phi_2\cr\Phi_3}=\pmatrix{\sin\chi(r)\,\cos{(\phi-\omega t)}\cr \sin\chi(r)\,\sin{(\phi-\omega t)}\cr\cos\chi(r)}\label{rotatngfld}\;.\ee 
The functions $A$, $B$ and $\chi$ are dimensionless, while $\Omega$ and $\omega$ have units of inverse-time. Without any loss of generality we can set $\chi(\infty)=0$.  Then for fields in the $n^{th}$ topological sector, $\chi(0)=n\pi$.
 Upon substituting (\ref{rotatngmtrc}) and (\ref{rotatngfld}) into the action (including the Gibbons-Hawking-York term) we get
\be S=\frac \pi{\kappa}\int\frac{ dtdr}{\sqrt{B}} \;\biggl\{ {\partial_r A}+\frac {r^3(\partial_r\Omega)^2}{2}+\frac{2 rB}{\ell^2} - { r\kappa}\biggl( A(\partial_r\chi)^2+\Bigl(\frac 1{r^2}-\frac{(\omega+\Omega)^2}{A}\Bigr){B}\sin^2\chi\biggr)\; \Biggr\}-S_{AdS}\;, \ee
where $\kappa=8\pi G$, and we set $\Lambda=-\frac 1{\ell^2}$. 
It is convenient to introduce the dimensionless radial variable $x=r/\ell$.    Then
\be S=\frac \pi{\kappa}\int\frac{ dtdx}{\sqrt{B}} \;\biggl\{ { A'}+\frac { x^3\tilde\Omega'^2}{2}+{2 xB} - {  x\kappa}\biggl( A\chi'^2+\Bigl(\frac 1{x^2}-\frac{ (\tilde \omega+\tilde\Omega)^2}{A}\Bigr){B}\sin^2\chi\biggr)\; \Biggr\}-S_{AdS}\;,\label{dmnslsactn} \ee
where $\tilde \Omega=\ell\,\Omega$ and $\tilde \omega=\ell\,\omega$, and the prime denotes a derivative with respect to $x$.
Upon extremizing the action with respect to variations in $A$, $B$, $\tilde \Omega$ and $\chi$, we get
\beqa \frac 12 (\ln B)'&=&\kappa\,x\Bigl(\chi'^2+\frac{ B}{A^2} (\tilde\omega+\tilde\Omega)^2\sin^2\chi \Bigr) \cr &&\cr
0&=& A' +\frac{x^3\tilde\Omega'^2}2  -2x B+ {\kappa\,x}\;\biggl(-A\chi'^2+\Bigl(\frac 1{x^2}-\frac{(\tilde\omega+\tilde\Omega)^2}{A}\Bigr){B}\sin^2\chi\biggr)\cr &&\cr
\left(\frac{x^3\tilde\Omega'}{\sqrt{B}}\right)'&=&  \frac{2\kappa x}A{\sqrt{B}}\,(\tilde\Omega+\tilde\omega)\,\sin ^2\chi
\cr &&\cr
\left(\frac {xA\chi'}{\sqrt{B}}\,\right)'&=&x\sqrt{B}\Bigl(\frac 1{x^2}-\frac{(\tilde\omega+\tilde\Omega)^2}{A}\Bigr)\sin\chi\cos\chi
\;,\label{eqsdmrnqdratic}\eeqa
respectively. 

Next we write down the solutions to (\ref{eqsdmrnqdratic}) in the asymptotic regions $x\rightarrow\infty$ and $x\rightarrow 0$.

\subsection{ $x\rightarrow\infty$}
For  the asymptotic region $x\rightarrow\infty$ we demand that $\chi\rightarrow 0$ and  that we recover anti-de Sitter space in the limit. 
  The large distance behavior for  $A$, $B$, $\tilde \Omega$ and $\chi$ can  be determined from (\ref{eqsdmrnqdratic}):
\beqa A &\rightarrow &{x^2}-M+\frac{J^2}{x^2}+\frac{\kappa\nu ^2  \left(\left(\tilde\Omega_\infty+ \tilde\omega\right)^2 +4 M+1\right) }{12x^4}  +{\cal O}\Bigl(\frac 1{x^6}\Bigr) \cr &&\cr  B &\rightarrow & 1-\frac{2\kappa \nu ^2}{x^4}+\frac{\kappa \nu ^2 \left(\left(\tilde\Omega_\infty+ \tilde \omega\right)^2 -8 M-2\right)}{3 x^6}+{\cal O}\Bigl(\frac 1{x^8}\Bigr) \cr &&\cr \tilde\Omega  &\rightarrow &\tilde\Omega_\infty+\frac{J}{x^2}+\frac{\kappa \nu ^2 \left(\tilde\Omega_\infty+ \tilde \omega-4 J \right)}{12\, x^6} +{\cal O}\Bigl(\frac 1{x^8}\Bigr)\cr &&\cr 
\chi  &\rightarrow &\frac{\nu }{x^2}+\frac{\nu   \left(-\left(\tilde\Omega_\infty+ \tilde \omega\right)^2+4 M+1\right)}{8\, x^4} +{\cal O}\Bigl(\frac 1{x^6}\Bigr)\;,\qquad\quad\;\; x\rightarrow\infty\;,\label{asmptcfrm}\eeqa
where $ M,\,J,\,\tilde\Omega_\infty$  and $\nu$ are constants, the first two being the mass and angular momentum parameters, respectively.  The solution is consistent with the standard large distance  behavior of the metric tensor for three-dimensional anti-de Sitter space with a localized matter source.\cite{Henneaux:2002wm}  The  Ricci scalar tends towards the $AdS^3$  value of $ -6$ in the limit.  The constant $\tilde\Omega_\infty$ can always be eliminated by transforming to the co-rotating frame at spatial infinity, where $\tilde\omega$ in the ansatz (\ref{rotatngfld}) gets replaced by $\tilde\Omega_\infty+ \tilde \omega$.  Conversely, we can transform to a frame where the $\sigma-$model fields are static by replacing  $\tilde\Omega_\infty$ by  $\tilde\Omega_\infty+ \tilde \omega$.

\subsection{$ x\rightarrow 0$}
Two possible power series expansions for $A$, $B$, $\tilde \Omega$ and $\chi$  exist near the origin.  Two of the functions, $A$ and $\tilde\Omega$, are singular at the origin for one solution, while all functions have a finite limit for the other.   For the former, $A,\tilde\Omega\sim \frac 1{x^2}$, as $x\rightarrow 0$. More specifically, near the origin the solution  has the form
\beqa A &\rightarrow &\frac{{J_0}^2}{x^2} \;-M_0\; +\; B_0 x^2\; -\;\frac \kappa 3  M_0\chi_2^2 {x^4}\;+\; {\cal O}({x^6}) \cr &&\cr
  B &\rightarrow &B_0\biggl(1\;+\;2\kappa\chi_2^2 x^4\;+\;\frac{ \kappa  \chi_2^2}{3{J_0}^2}(B_0+8M_0)  x^6\;+\;{\cal O}({x^8})\biggr)\cr &&\cr
 \tilde\Omega  &\rightarrow &\frac{{J_0}}{x^2} \;+\;\tilde\Omega_0\;-\kappa {J_0}  \chi_2^2 x^2 \;-\;\frac {2\kappa M_0 \chi_2^2 }{3 {J_0}} x^4\;+\;{\cal O}({x^6}) \cr &&\cr 
\chi  &\rightarrow &n\pi\;+\;\chi_2 x^2\;+\;\frac{\chi_2M_0}{2{J_0}^2 }x^4\; +\;{\cal O}({x^6})\;,\qquad\quad\;\; x\rightarrow 0\;,\label{nrrignfrm}\eeqa
where ${J_0}$, $M_0$, $B_0$,  $\tilde\Omega_0$ and $\chi_2$ are constants.
For finite ${J_0}\ne 0$, $M_0$, $B_0$,  $\tilde\Omega_0$,
 the time-time component  $g_{tt}=-A+x^2\tilde\Omega^2$, as well as the remaining components of the metric tensor, are bounded at the origin, 
\be ds^2 \sim (M_0 + 2{J_0}\tilde\Omega_0)\,dt^2 +\frac{B_0}{J_0^2}  x^2\, dx^2 + 2 {J_0} \,dtd\phi + x^2 \,d\phi^2\label{mtrcnrorgn}\ee
All scalars constructed from the curvature tensor are bounded in  the $x\rightarrow 0$ limit, e.g. the  Ricci scalar tends toward $\frac {4\kappa{J_0}^2 \chi_2^2}{B_0} -6$. Nevertheless, a  causal  singularity exists at the origin for this solution.   The metric tensor near the origin closely resembles  that of the BTZ black hole.\cite{Banados:1992gq}. For the numerical solutions discussed in section 3 there are no horizons at finite $x$, and so  the singularity  is naked for all such  solutions.

The  power series solution (\ref{mtrcnrorgn}) is not valid for $J_0=0$. For this case one has the alternative power series solution
\beqa A &\rightarrow &-M_0\; -\; M_0 x^2\; +\;\frac \kappa 8 \chi_1^2 \Bigl(-3M_0+(\tilde\Omega_0+\tilde\omega)^2\Bigr) {x^4}\;+\; {\cal O}({x^6}) \cr &&\cr
  B &\rightarrow &-M_0\;-\;\kappa M_0\chi_1^2 x^2+\;
\frac\kappa 8\chi_1^2\biggl(-M_0\Bigl(2\chi_1^2(5\kappa -1)-9 \Bigr)+(\tilde\Omega_0+\tilde\omega)^2\biggr) x^4\;+\;{\cal O}({x^6})\biggr)\cr &&\cr
 \tilde\Omega  &\rightarrow &\tilde\Omega_0\;+\frac\kappa 4  \chi_1^2(\tilde\Omega_0+\tilde\omega) x^2 \;-\;\frac {\kappa}{48 M_0} \chi_1^2 (\tilde\Omega_0+\tilde\omega) \biggl(-M_0\Bigl(2\chi_1^2(4\kappa -1)-7 \Bigr)-(\tilde\Omega_0+\tilde\omega)^2\biggr) x^4\;+\;{\cal O}({x^6}) \cr &&\cr 
\chi  &\rightarrow &n\pi\;+\;\chi_1 x\;-\;\frac{\chi_1}{24 M_0}\biggl(-M_0\Bigl(2\chi_1^2(3\kappa -1)-9 \Bigr)-3(\tilde\Omega_0+\tilde\omega)^2\biggr)x^3\; +\;{\cal O}({x^5})\;,\qquad\quad\;\; x\rightarrow 0\;,\cr&&\label{fntnrrign}\eeqa
where all functions have a finite limit.  This solution is parametrized by  $M_0$,  $\tilde\Omega_0$ and $\chi_1$.
The invariant length near the origin  takes the form
\be ds^2\sim M_0(1+x^2) dt^2 +\Bigl(1+(\kappa\chi_1^2-1)x^2\Bigr) dx^2+ x^2(d\phi+\tilde\Omega_0 dt)^2 \;\ee
For a Lorentzian space-time near the origin we need that $M_0<0$.  When $\tilde\Omega_0=0$, any $t-$slice  approaches flat  Euclidean space as $x\rightarrow 0$. When $\tilde\Omega_0\ne 0$, the space-time near the origin is rotating.  In either case the space-time is singularity free.

\section{Numerical solutions}

We have not found any analytic solutions to  (\ref{eqsdmrnqdratic})  away from the asymptotic regions and therefore rely on numerical methods.
We  numerically integrate (\ref{eqsdmrnqdratic}) subject to the asymptotic expressions (\ref{asmptcfrm}) near the $AdS^3$ boundary to obtain $A$, $B$, $\tilde \Omega$ and $\chi$ at finite $x$.
For   topological solitons  $\chi(0)$ must be an integer multiple of $\pi$. The topological solitons solutions can be parametrized by  the constants $J, M$, $\nu$ and  $\tilde\Omega_\infty$ appearing in (\ref{asmptcfrm}). One strategy for obtaining solutions is to first fix  three of the parameters (along with $\kappa$ and $\tilde\omega$), and then apply shooting methods to  tune the remaining one   such that $\chi\rightarrow n\pi$ as $ x\rightarrow 0$, where the winding number $n$ is equal to a nonzero integer.  Near the origin the solutions must satisfy either (\ref{nrrignfrm}) or (\ref{fntnrrign}), corresponding to type $i)$ or $ii)$ solutions, respectively.
The parameters appearing in  (\ref{nrrignfrm}) or (\ref{fntnrrign}) 
  can then be determined numerically from $ M,\,J$, $\nu$  and  $\tilde\Omega_\infty$.  Conversely, given  ${J_0}$, $M_0$, $B_0$,  $\tilde\Omega_0$ and $\chi_2$ of the expression (\ref{nrrignfrm})  or  $M_0$,   $\tilde\Omega_0$ and $\chi_1$ of  (\ref{fntnrrign}) (and the winding number $n$) we can numerically determine the  parameters $ M,\,J$, $\nu$ and  $\tilde\Omega_\infty$  describing the large distance behavior.

\subsection{Type $i)$ solutions}

Numerical solutions  satisfying (\ref{nrrignfrm}) near the origin corresponding to type $i)$ solutions are found for  large regions of the parameter space.  Examples of the behavior of the functions $A(x)$ and $\chi(x)$  for these solutions appear in figures 1 through 5.  There we plot $\chi$ versus log$(x)$ and log$(A)$ versus log$(x)$  for different    values of $n$, $\tilde\omega$, $M$, $J$ and  $\kappa$. We set $\tilde\Omega_\infty=0$, which means that we are working in the co-rotating frame as $x\rightarrow\infty$.  In the captions we  list the  fitted values for $\nu$ for each solution. 
One example which appears in all  figures 1-5 is an $n=1$ soliton with $\kappa=\tilde\omega=M=J=1$  and $\nu\approx 2.33 $.   For this example the  functions $A$, $B$, $\tilde \Omega$ and $\chi$ tend towards (\ref{nrrignfrm}) as $x\rightarrow 0$, with the following values for the short distance parameters:  $M_0\sim -.34$, $J_0\sim .094$, $B_0\sim .0035$,  $\tilde\Omega_0\sim 1.92$ and $\chi_2\sim-14.9$.

Solutions  $i)$ with winding number one, two and three  with $M=J=\kappa=\tilde\omega=1$ are shown in figures 1a and 1b. $n=1$ solutions are plotted for  different rotation velocities $\tilde\omega$, including zero, (with $M=J=\kappa=1$) in figures 2a and 2b.    $n=1$ solutions are plotted for  different values of the mass parameter, including $0$ and $-1$, (with $J=\kappa=\tilde\omega=1$)  in figures 3a and 3b   and  different values of the angular momentum parameter, including $0$, (with $M=\kappa=\tilde\omega=1$) in figures 4a and 4b. Finally, $n=1$ solutions are plotted in figures 5a and 5b for  different values of $\kappa$ (with $M=J=\tilde\omega=1$). From the results in figures 2 and 4, neither  a nonzero rotation velocity $\tilde\omega$ in the internal space nor a nonzero angular momentum $J$ is necessary to stabilize the soliton, since we find solutions when either  $\tilde\omega$ or $J$ are zero. On the other hand, we find no  solutions when both $\tilde\omega$ and $J$ vanish, which is consistent with the no-go result in \cite{Bizon:2004wa}.  In addition, we find novel  solutions where both the mass and angular momentum parameters vanish,  $M=J=0$, and one with $M=-1,\;J=0$.  If one takes these as  parameters for the   BTZ black hole, the  former would correspond to a zero mass black hole and the latter would correspond to  anti-de Sitter space.  An example  of  a soliton with $M=J=0$  occurs for  $\kappa=\tilde\omega=1$, $\nu\approx .77$, and a soliton with $M=-1,\;J=0$ occurs for $\kappa=\tilde\omega=1$, $\nu\approx 1.061$.  
As required, $A$ approaches $x^2$ as $x\rightarrow \infty$ for all of the above examples, while it  does not pass through zero for any $x$.  The latter behavior  indicates that there are no horizons. $A$ and $\Omega$ go as $1/x^2$ near the origin.
\begin{figure}[placement h]
\centering
\begin{subfigure}{.45\textwidth}
  \centering
  \includegraphics[height=1.7in,width=2.5in,angle=0]{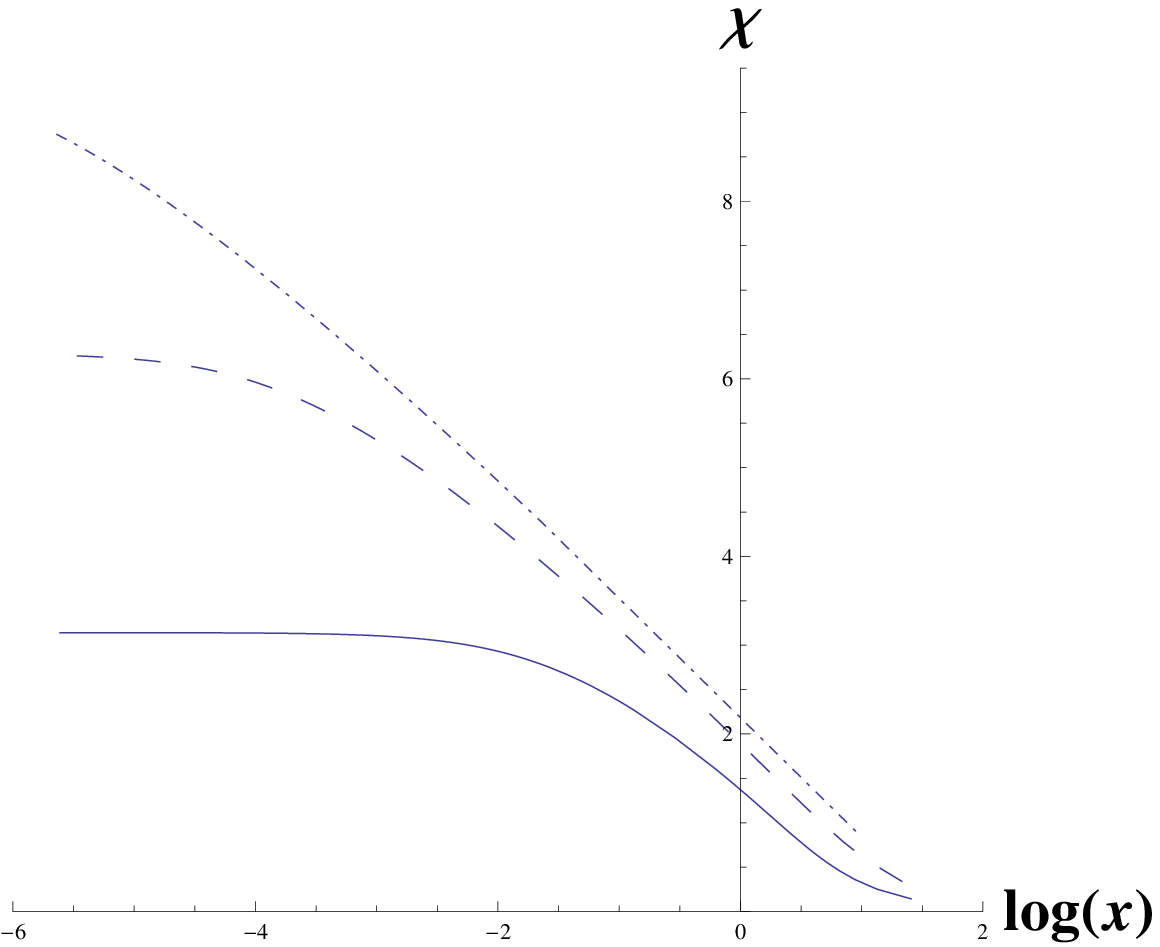}
  \caption{$\chi$ vs. $\log{x}$}
  \label{fig:sub1}
\end{subfigure}%
\begin{subfigure}{.45\textwidth}
  \centering
  \includegraphics[height=1.7in,width=2.5in,angle=0]{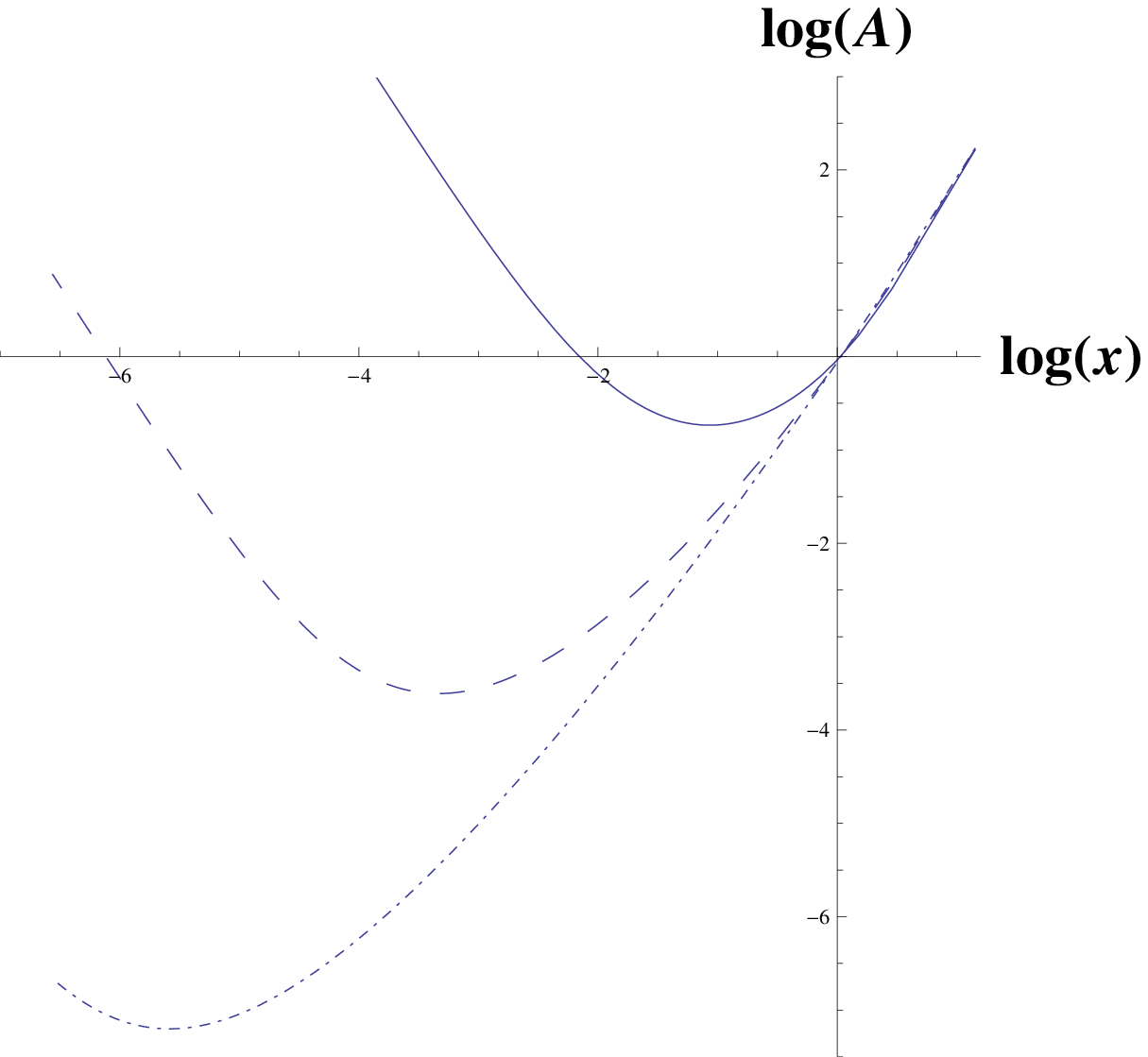}
  \caption{$\log{A}$ vs. $\log{x}$}
  \label{fig:sub1}
\end{subfigure}%
\caption {{\bf Varying $n$.}  Self-gravitating rotating solitons with  parameters:  $M=J=\kappa=\tilde\omega=1$.  $\chi $ versus $\log{x}$ is plotted in figure (a) and $\log{ A}$ versus $\log{x}$ is plotted in figure (b).  $\nu\approx 2.33 $ gives the $n=1$ soliton (solid curve),  $\nu\approx 4.92 $ gives the $n=2$ soliton (dashed  curve) and  $\nu\approx 7.52$ gives the $n=3$ soliton (dot-dashed curve curve).}  
\label{fig:test}
\end{figure}
\begin{figure}[placement h]
\centering
\begin{subfigure}{.45\textwidth}
  \centering
  \includegraphics[height=1.7in,width=2.5in,angle=0]{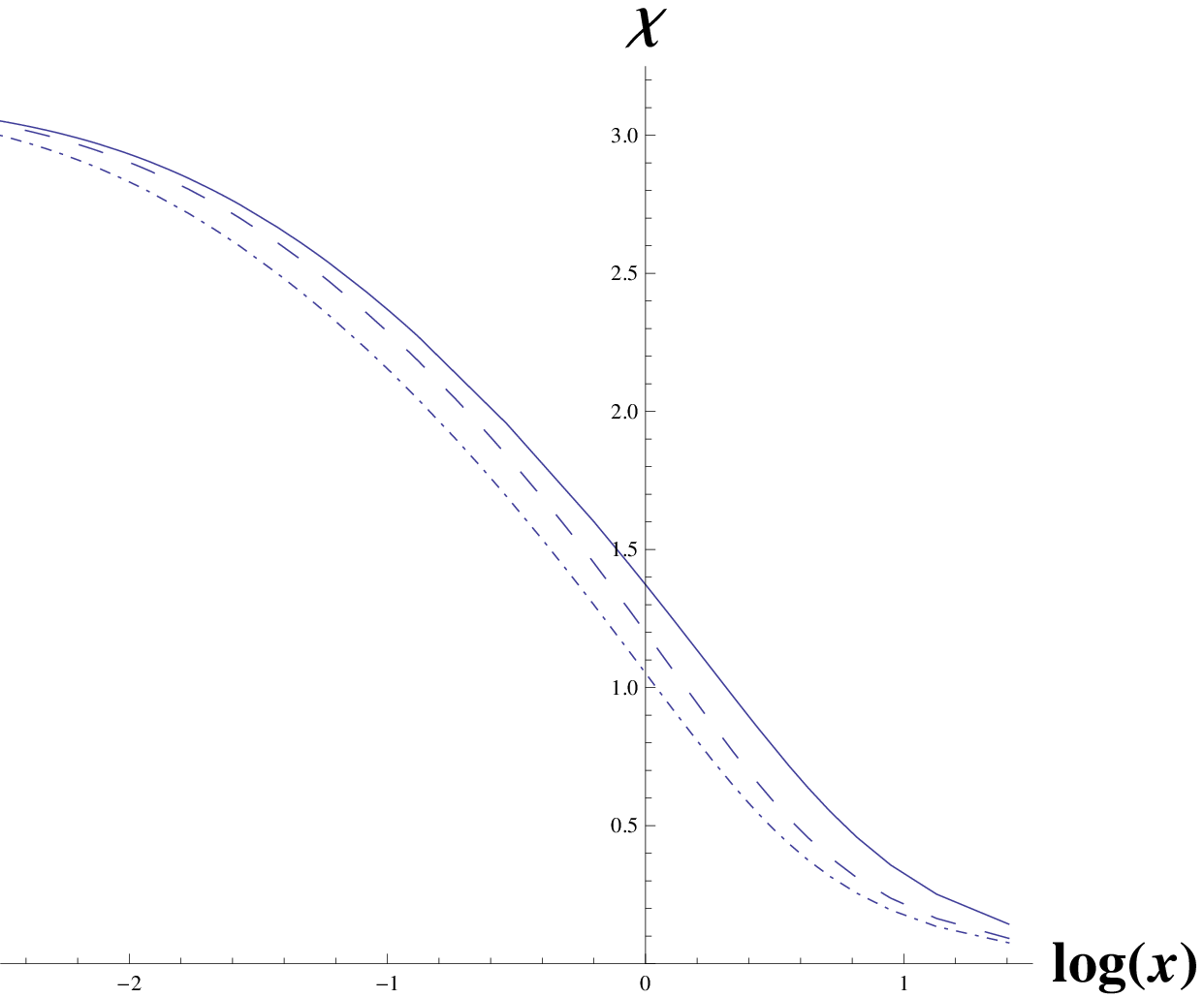}
  \caption{$\chi$ vs. $\log{x}$}
  \label{fig:sub1}
\end{subfigure}%
\begin{subfigure}{.45\textwidth}
  \centering
  \includegraphics[height=1.7in,width=2.5in,angle=0]{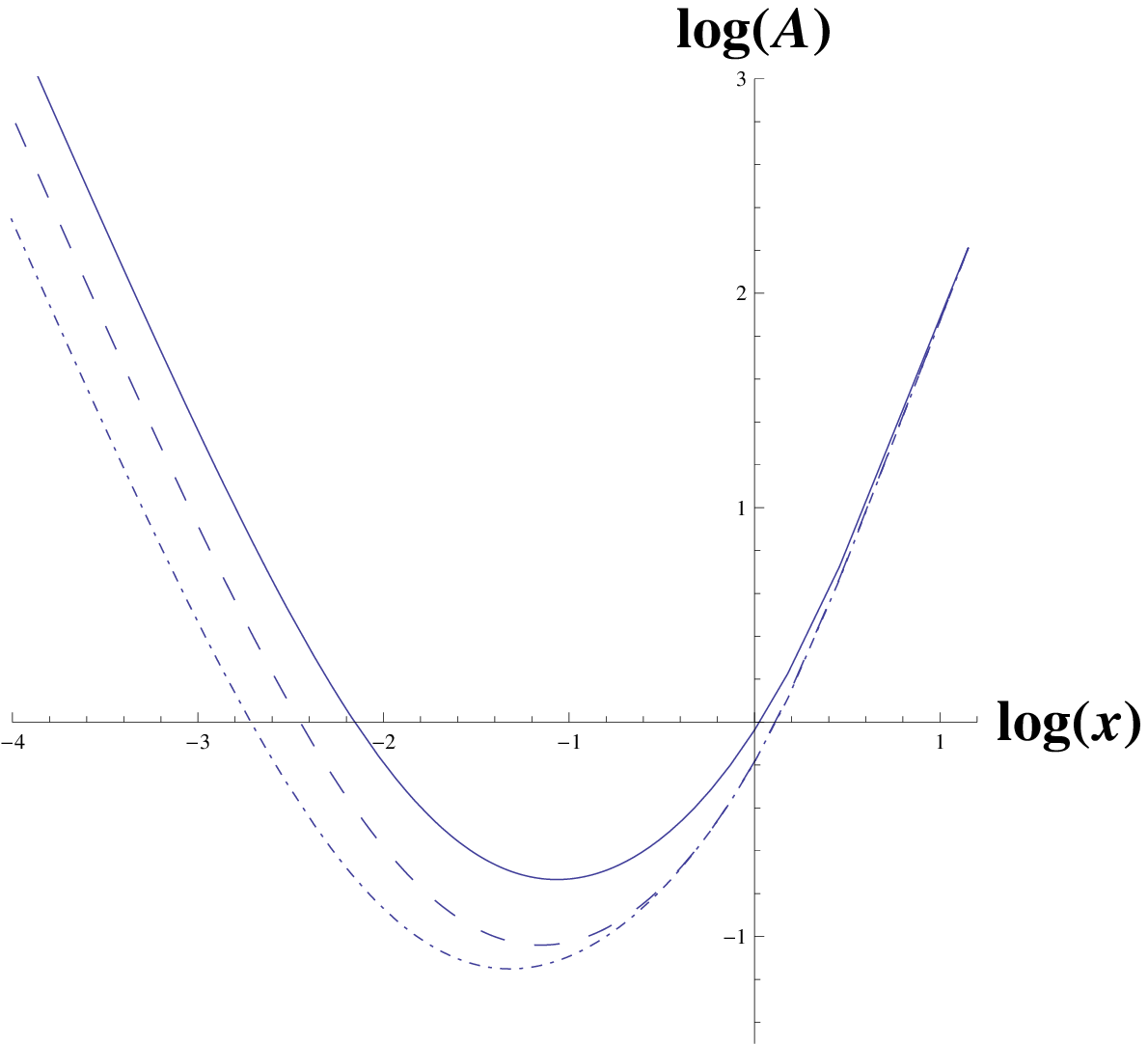}
  \caption{$\log{A}$ vs. $\log{x}$}
  \label{fig:sub1}
\end{subfigure}%
\caption {{\bf Varying  $\tilde\omega$.} $n=1$  self-gravitating solitons with  parameters:  $M=J=\kappa=1$ for different values of $\tilde\omega$.  $\chi $ versus $\log{x}$ is plotted in figure (a) and $\log{ A}$ versus $\log{x}$ is plotted in figure (b).  $\nu\approx 2.33 $ gives the soliton with $\tilde\omega=1$  (solid curve),  $\nu\approx 1.48 $ gives the  soliton with $\tilde\omega=0$ (dashed  curve) and  $\nu\approx 1.22$ gives the soliton with $\tilde\omega=-1$  (dot-dashed curve).}  
\label{fig:test}
\end{figure}
\begin{figure}[placement h]
\centering
\begin{subfigure}{.45\textwidth}
  \centering
  \includegraphics[height=1.7in,width=2.5in,angle=0]{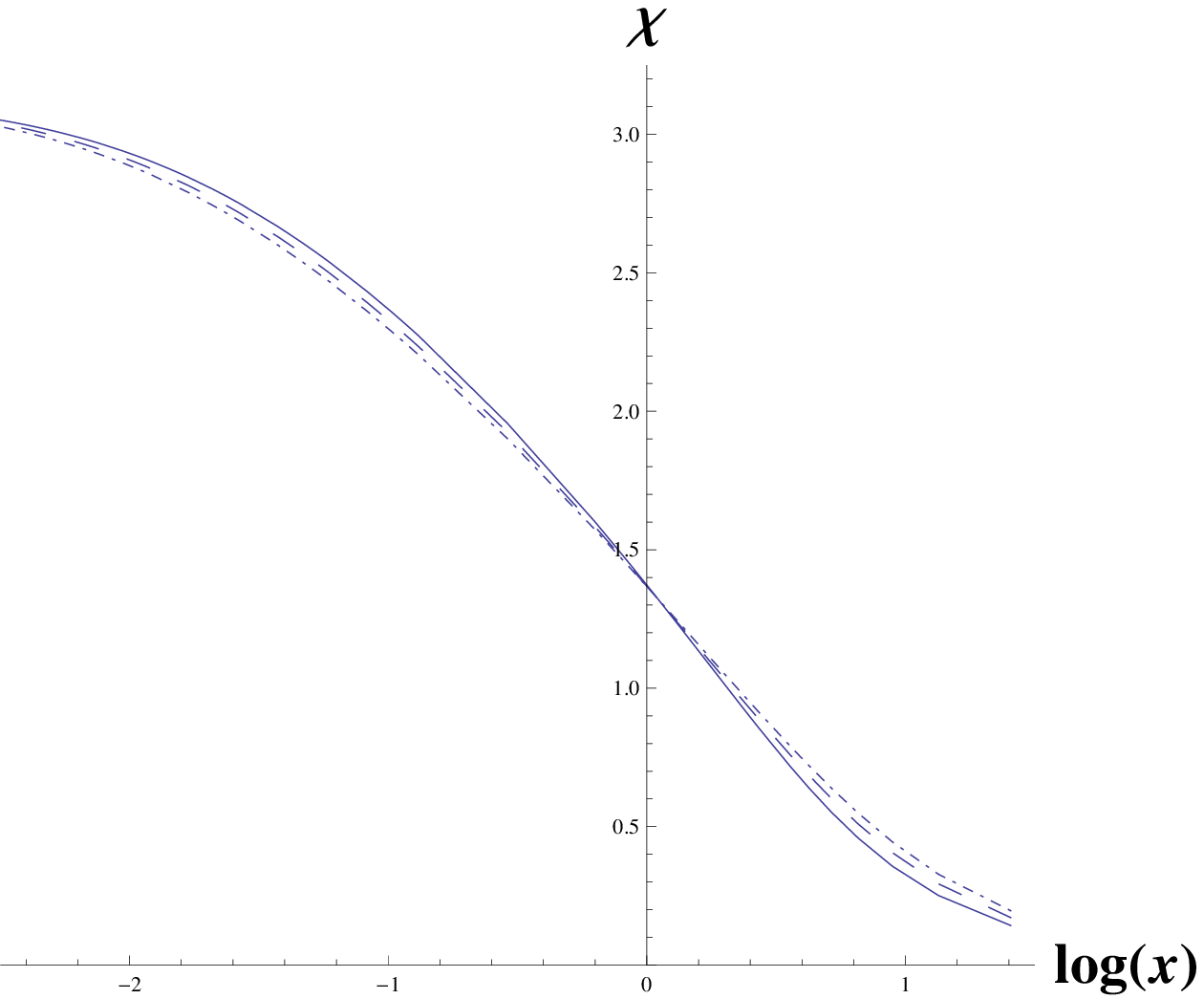}
  \caption{$\chi$ vs. $\log{x}$}
  \label{fig:sub1}
\end{subfigure}%
\begin{subfigure}{.45\textwidth}
  \centering
  \includegraphics[height=1.7in,width=2.5in,angle=0]{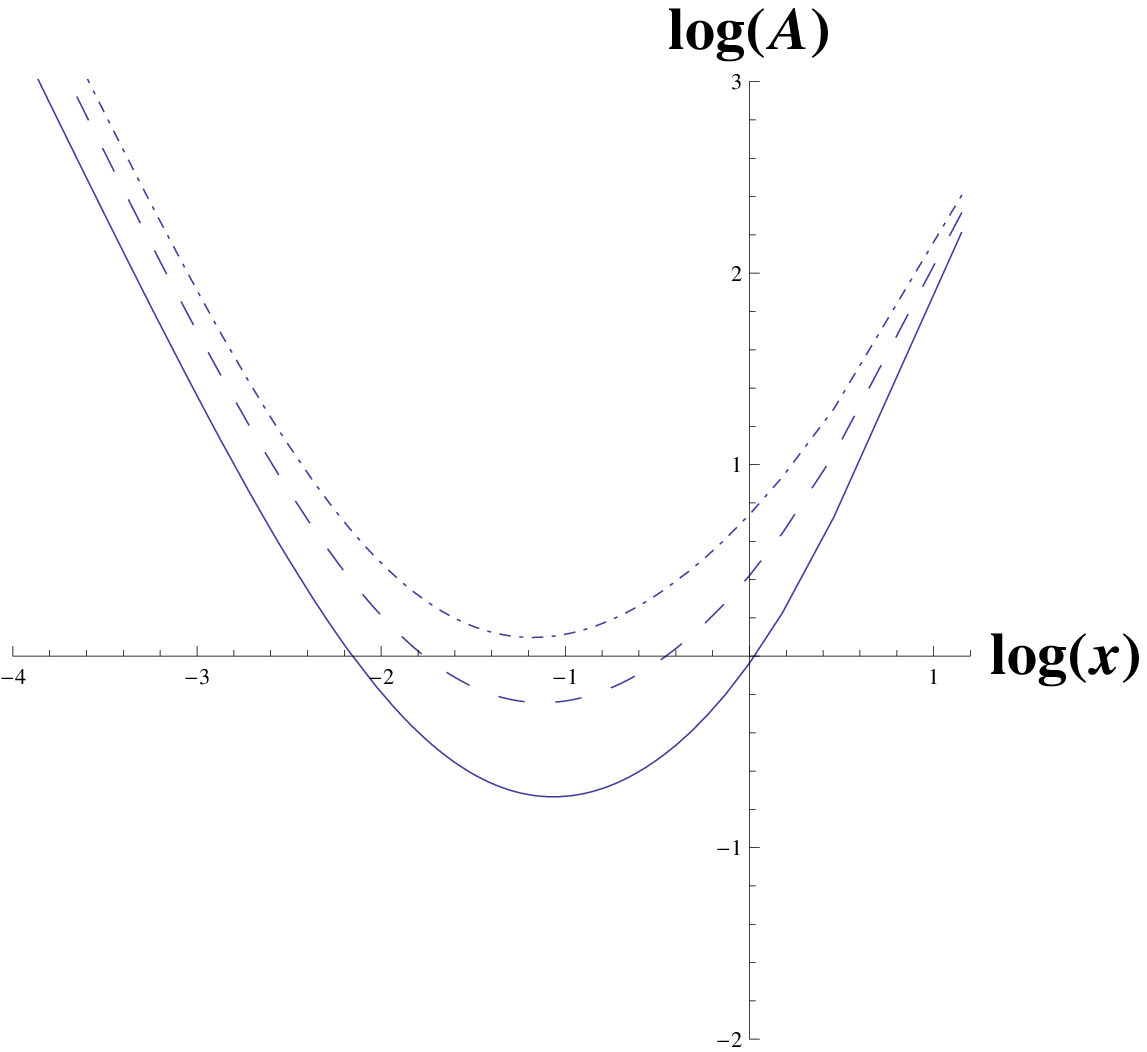}
  \caption{$\log{A}$ vs. $\log{x}$}
  \label{fig:sub1}
\end{subfigure}%
\caption {{\bf Varying  $M$.} $n=1$  self-gravitating solitons with  parameters:  $J=\kappa=\tilde\omega=1$ for different values of  the mass parameter $M$.  $\chi $ versus $\log{x}$ is plotted in figure (a) and $\log{ A}$ versus $\log{x}$ is plotted in figure (b).  $\nu\approx 2.33 $ gives the soliton with $M=1$  (solid curve),  $\nu\approx 2.89 $ gives the  soliton with $M=0$ (dashed  curve) and  $\nu\approx 3.42$ gives the soliton with $M=-1$  (dot-dashed curve).}  
\label{fig:test}
\end{figure}
\begin{figure}[placement h]
\centering
\begin{subfigure}{.45\textwidth}
  \centering
  \includegraphics[height=1.7in,width=2.5in,angle=0]{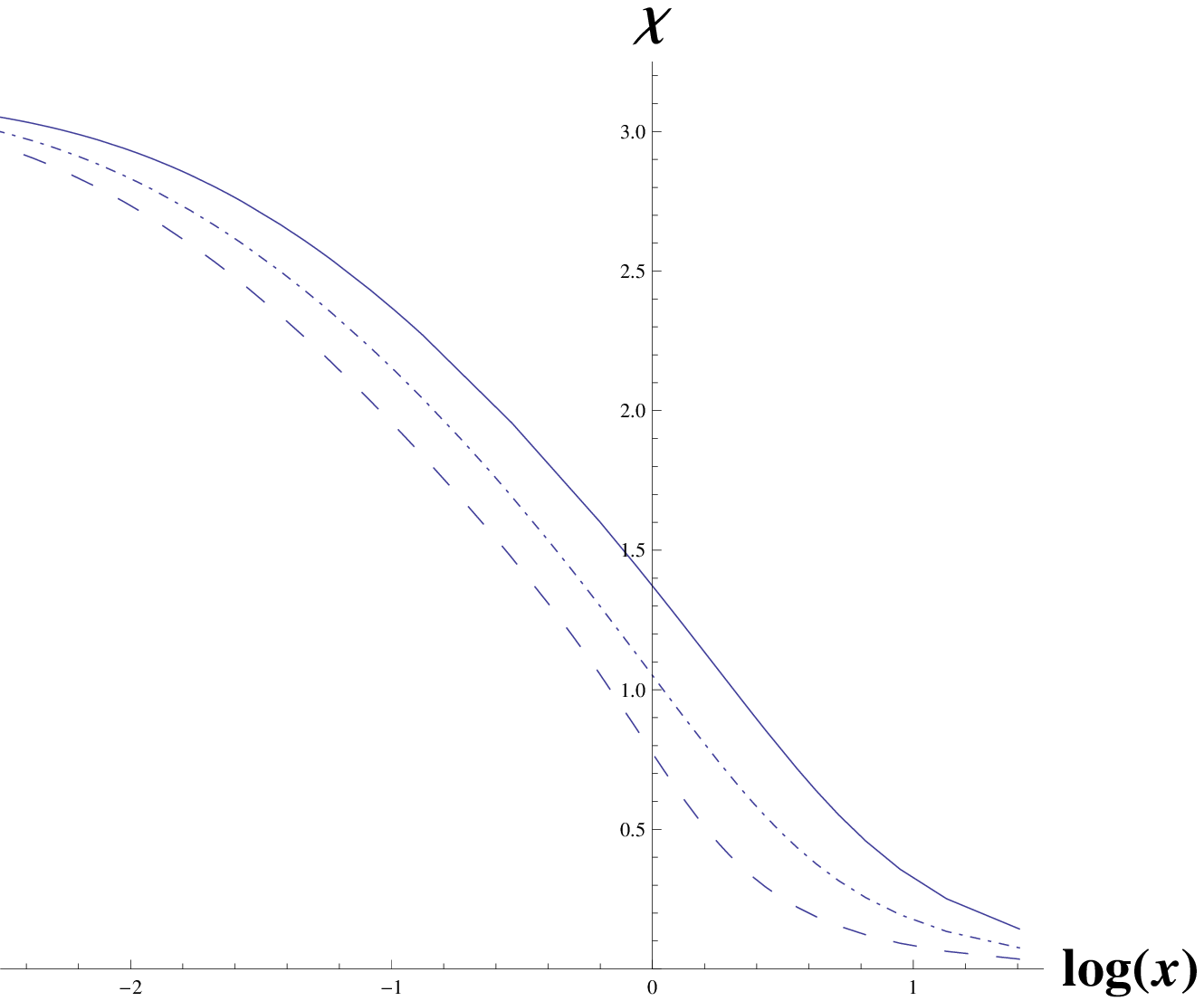}
  \caption{$\chi$ vs. $\log{x}$}
  \label{fig:sub1}
\end{subfigure}%
\begin{subfigure}{.45\textwidth}
  \centering
  \includegraphics[height=1.7in,width=2.5in,angle=0]{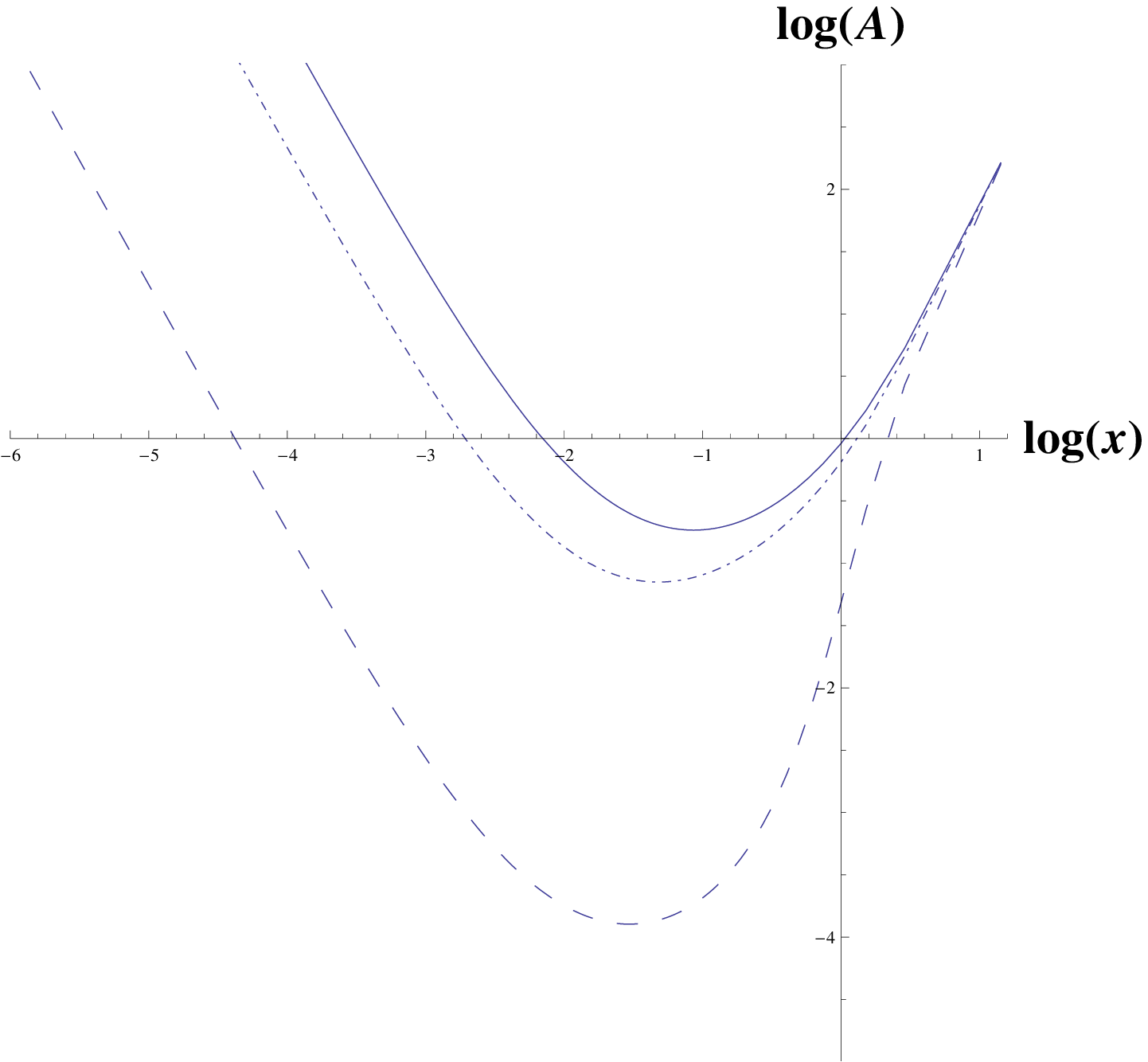}
  \caption{$\log{A}$ vs. $\log{x}$}
  \label{fig:sub1}
\end{subfigure}%
\caption {{\bf Varying  $J$.} $n=1$  self-gravitating solitons with  parameters:  $M=\kappa=\tilde\omega=1$ for different values of  the angular momentum parameter $J$.  $\chi $ versus $\log{x}$ is plotted in figure (a) and $\log{ A}$ versus $\log{x}$ is plotted in figure (b).  $\nu\approx 2.33 $ gives the soliton with $J=1$  (solid curve),  $\nu\approx .569 $ gives the  soliton with $J=0$ (dashed  curve) and  $\nu\approx 1.22$ gives the soliton with $J=-1$  (dot-dashed curve).}  
\label{fig:test}
\end{figure}
\begin{figure}[placement h]
\centering
\begin{subfigure}{.45\textwidth}
  \centering
  \includegraphics[height=1.7in,width=2.5in,angle=0]{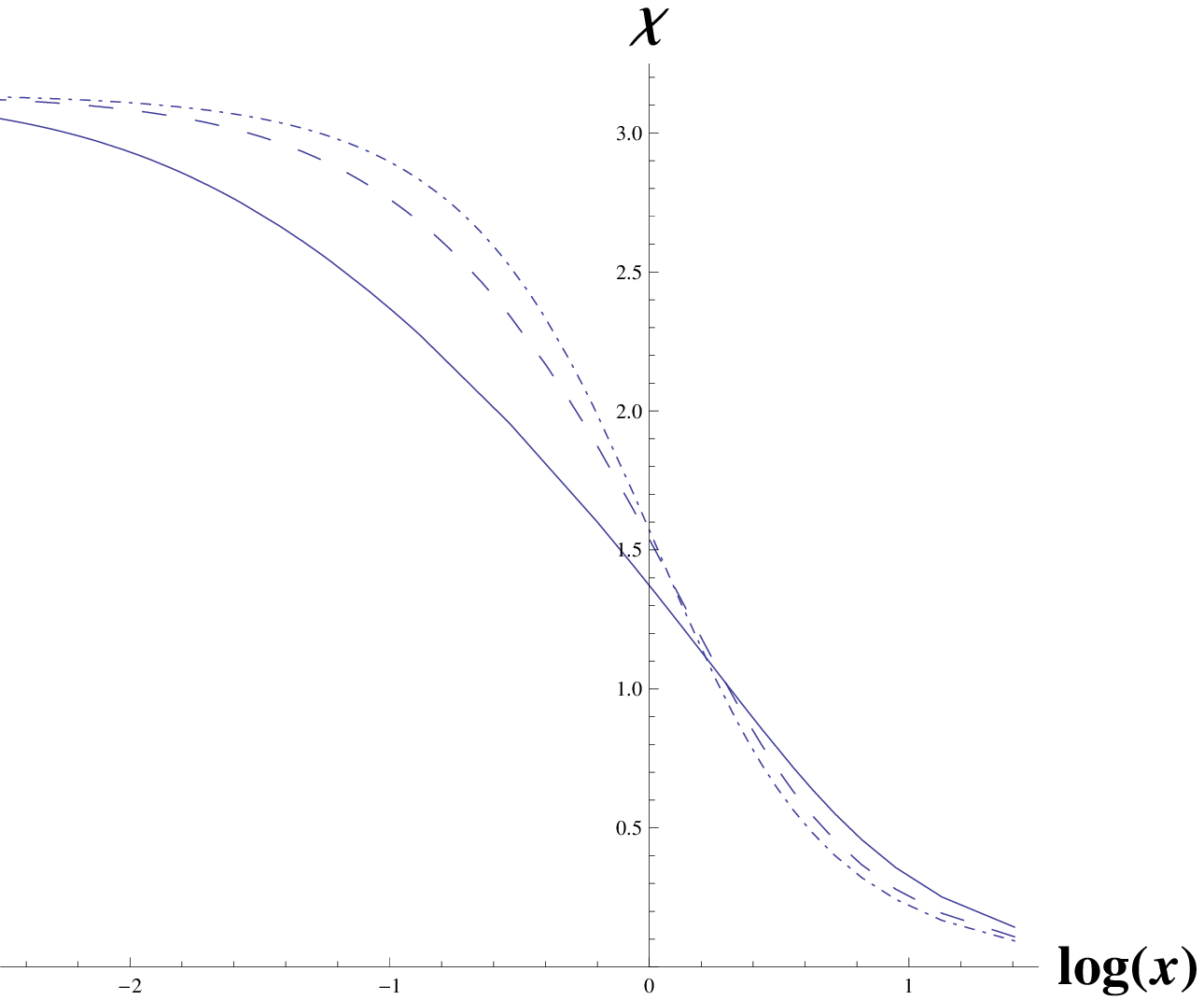}
  \caption{$\chi$ vs. $\log{x}$}
  \label{fig:sub1}
\end{subfigure}%
\begin{subfigure}{.45\textwidth}
  \centering
  \includegraphics[height=1.7in,width=2.5in,angle=0]{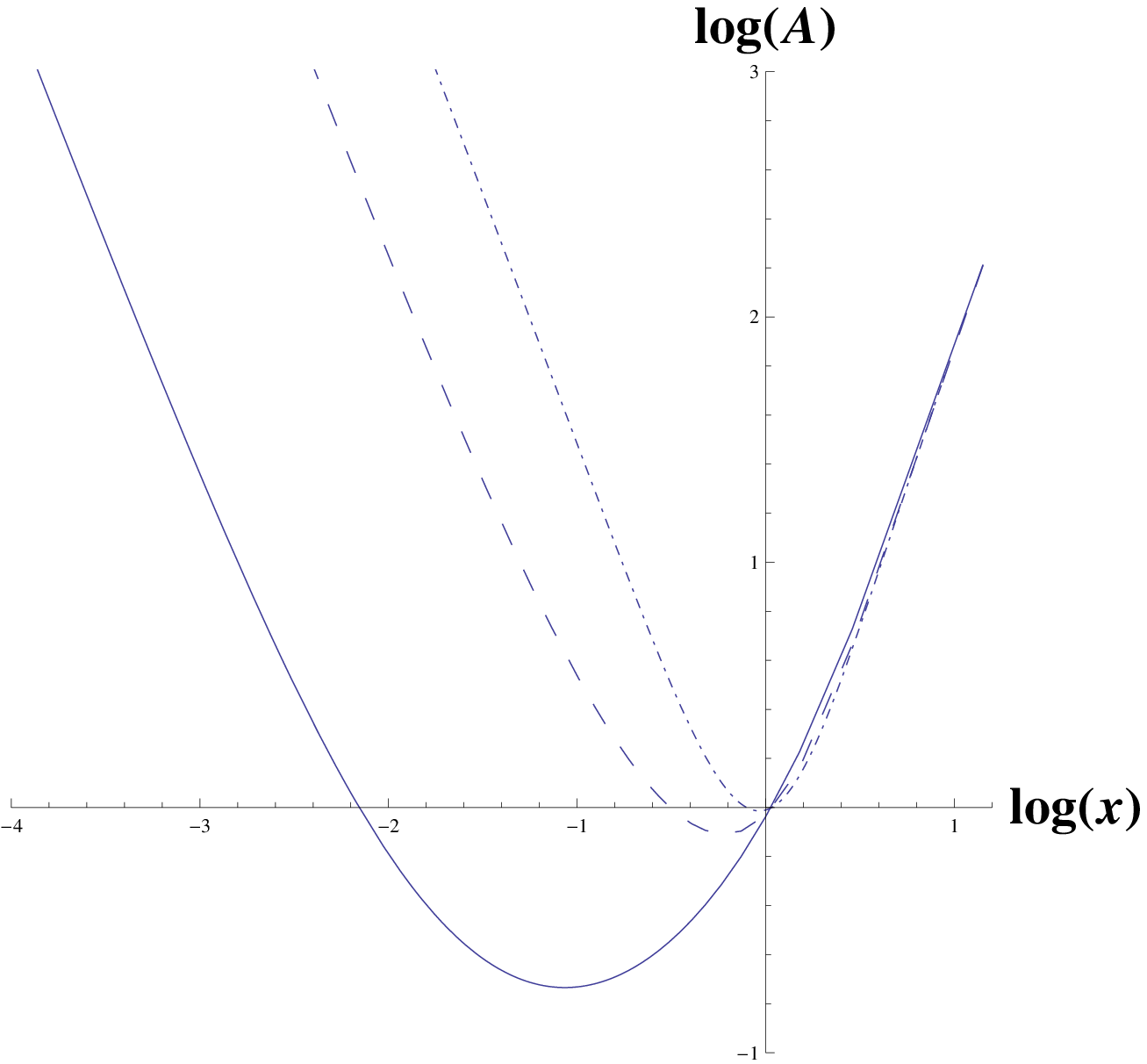}
  \caption{$\log{A}$ vs. $\log{x}$}
  \label{fig:sub1}
\end{subfigure}%
\caption {{\bf Varying  $\kappa$.} $n=1$  self-gravitating solitons with  parameters:  $M=J=\tilde\omega=1$ for different values of  $\kappa$.  $\chi $ versus $\log{x}$ is plotted in figure (a) and $\log{ A}$ versus $\log{x}$ is plotted in figure (b).  $\nu\approx 2.33 $ gives the soliton with $\kappa=1$  (solid curve),  $\nu\approx 1.76 $ gives the  soliton with $\kappa=.25$ (dashed  curve) and  $\nu\approx 1.52$ gives the soliton with $\kappa=.05$  (dot-dashed curve).}  
\label{fig:test}
\end{figure}

\subsection{Type $ii)$ solutions}

 Numerical solutions $ii)$ can also be found for which all of the functions are bounded,  including at the origin where they approach (\ref{fntnrrign}).  These solutions cover a smaller region in parameter space than $i)$ since they correspond to the limiting case of $J_0\rightarrow 0$ in  (\ref{mtrcnrorgn}).  
By integrating from $x\rightarrow \infty$ using  (\ref{asmptcfrm})  and from $x\rightarrow 0$ using  (\ref{fntnrrign}) we can match the four functions $A$, $B$, $\Omega$ and $\chi$, along with their derivatives, at  finite $x$ to arbitrary accuracy. An example of such a solution is  shown in figure 6(a), where  $M_0\approx -.193$, $\tilde\Omega_0\approx -2.09$, $\chi_1\approx -8.08$, $\omega\approx 1.37$ and $\kappa\approx .246$.  All four functions are well behaved at the origin.  By matching the functions and their derivatives at $x=1$, we determined the parameters of the large distance solution  (\ref{asmptcfrm}) to be  $M\approx .0539$, $J\approx .0817$, $\Omega_\infty\approx -2.77$ and $\nu\approx .241$.    As  in the previous figures, $A$ does not cross the $x-$axis, indicating no horizons.  The function $\chi(x)$ monotonically decreases from  $\pi$ to $0$ as $x$ goes from $0$ to $\infty$.  On the other hand, solutions can also be found where $\chi(x)$ has multiple nodes, as is illustrated in figure  6(b) for solutions with zero, one and two nodes.  The depicted $n=1$ solitions have common values for $\kappa$, $\omega$ and $M_0$, and differing values of the parameters $\chi_1$ and $\tilde \Omega_0$.  

The space of nonsingular solutions can be parametrized by $\omega$, $\kappa$, $M_0$, $\tilde\Omega_0$ and $\chi_1$.  We span the parameter space for the zero, one and two node solutions in figures 7, 8(a) and 8(b).  Keeping $\kappa$ and $\omega$ fixed, we plot $-\chi_1$ versus $\tilde\Omega_0$ in figure 7.   $\chi_1(\tilde\Omega_0)$ is seen to be multi-valued, with a cusp singularity occuring for the zero node solutions at some minimum value of  $\tilde\Omega_0$.  (Analogous behavior has been noted for self-gravitating Skyrmions.\cite{Bizon:1992gb})  $-\chi_1(\kappa)$ [with $\omega$ and $M_0$ held fixed] is plotted in figure 8(a) and $-\chi_1(-M_0)$ [with $\omega$ and $\kappa$ held fixed] is plotted in figure 8(b).  Finite domains are seen for both of these functions, implying upper bounds on the allowed values for $\kappa$ and $-M_0$.  We also get no soliton solutions in the limiting cases of $\kappa\rightarrow 0$ and $-M_0\rightarrow 0$. [The latter limit corresponds to the function $A$ vanishing at the origin, indicating a horizon in the zero radius limit.  So if solutions existed with $-M_0\rightarrow 0$ they  could coincide with the zero radius limit of  the horizon of a hairy black hole.
Thus the absence of such solutions  is consistent with not finding any black hole solutions with $\sigma-$model hair, which is what we report in  section five.] 
\begin{figure}[placement h]
\centering
\begin{subfigure}{.5\textwidth}
  \centering
  \includegraphics[height=2.15in,width=2.5in,angle=0]{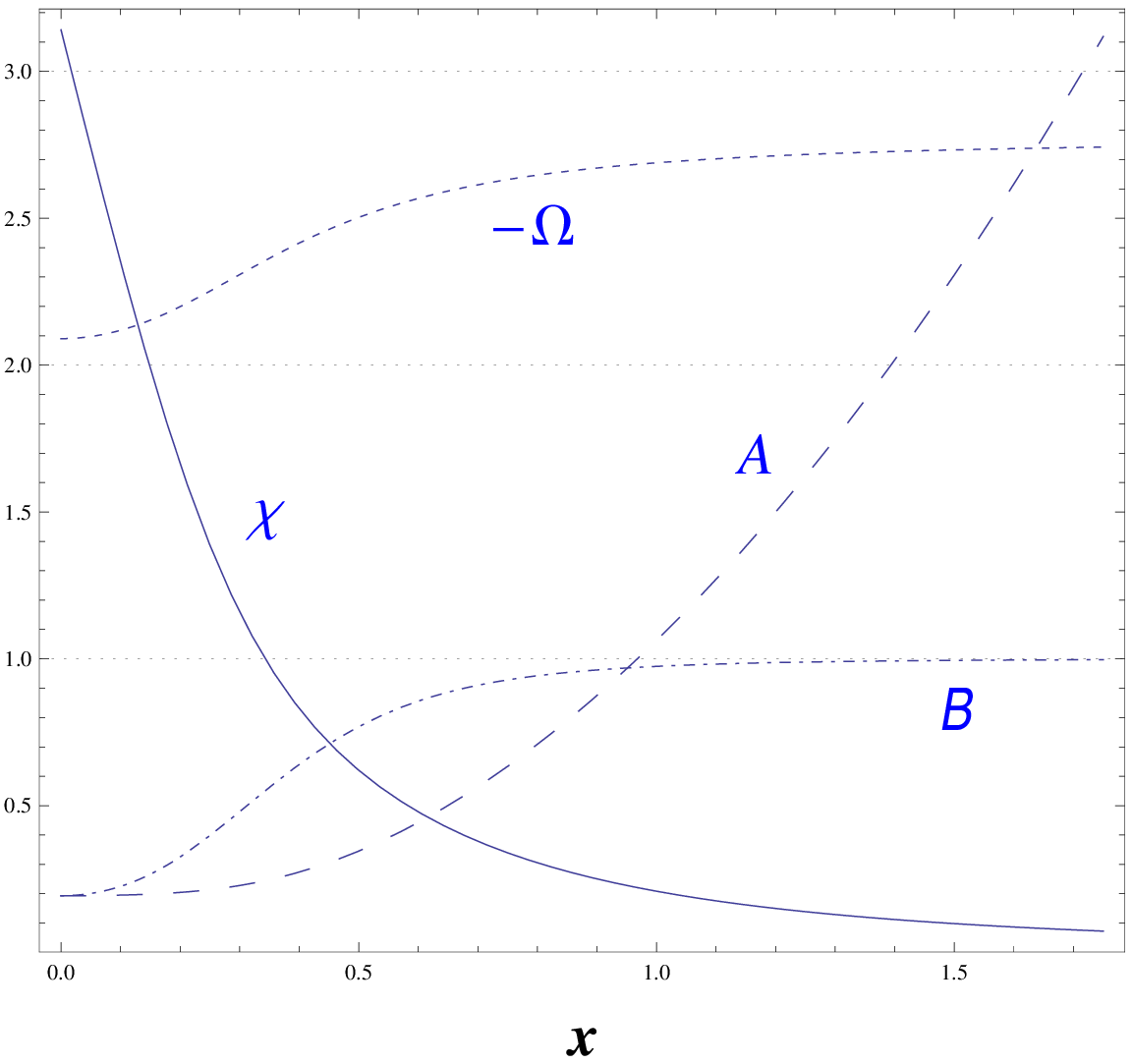}
  \caption{}
  \label{fig:sub1}
\end{subfigure}%
\begin{subfigure}{.5\textwidth}
  \centering
  \includegraphics[height=2.15in,width=2.5in,angle=0]{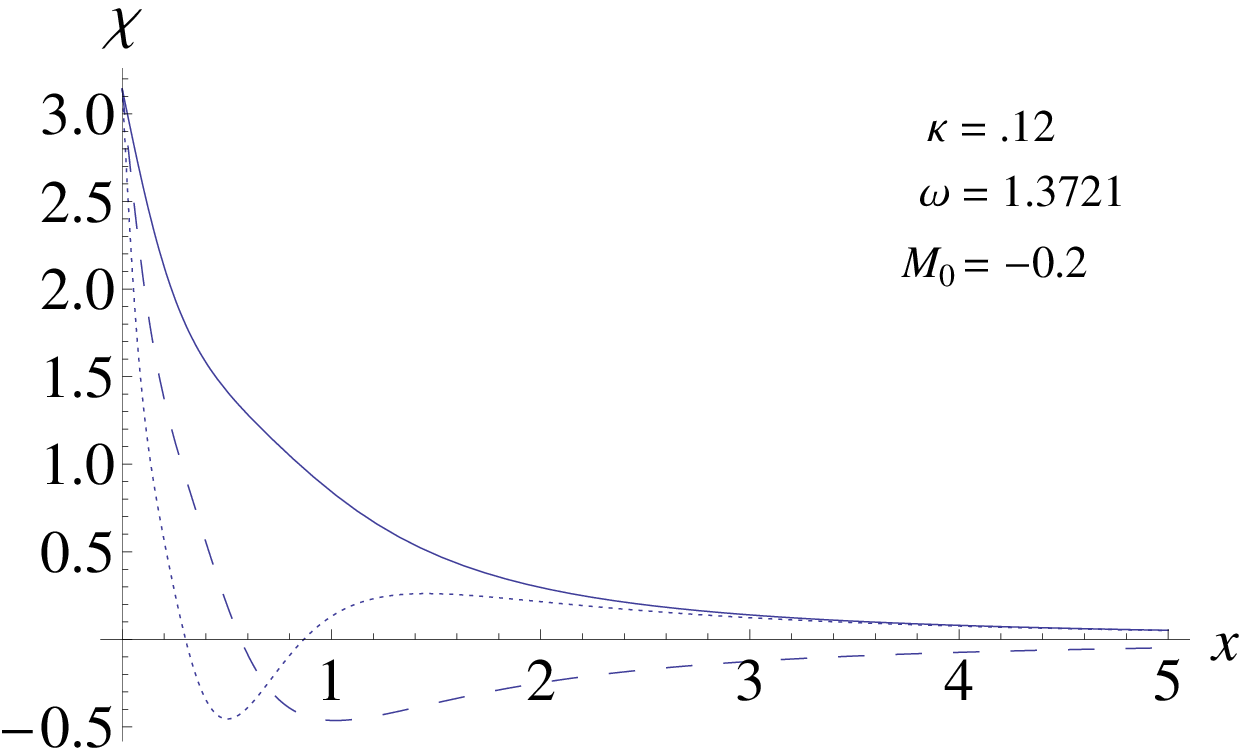}
  \caption{}
  \label{fig:sub1}
\end{subfigure}%
\caption {{\bf Singularity-free solutions}.  In (a), $\chi$ (solid curve), $A$ (large dashed curve),  $B$  (dot-dashed curve)  and $-\tilde\Omega$ (small dashed curve) are plotted versus $x$ for a singularity-free solution for the following values of the parameters introduced in (\ref{fntnrrign}): $M_0\approx -.193$, $\tilde\Omega_0\approx -2.09$, $\chi_1\approx -8.08$, $\omega\approx 1.37$ and $\kappa\approx .246$.  In (b),  $\chi$(x) is plotted for three different soliton solutions having $\kappa\approx.12,$ $\omega\approx 1.3721$ and $M_0\approx-.2$ The monotonically decreasing solution (solid curve) has  $\chi_1\approx -5.80$ and $\tilde\Omega_0\approx -3.01$; The single node solution (dashed curve) has  $\chi_1\approx -11.94$ and $\tilde\Omega_0\approx -3.52$; The double node solution (dotted curve) has  $\chi_1\approx -23.25$ and $\tilde\Omega_0\approx -4.19$.}  
\label{fig:test}
\end{figure}
\begin{figure}[placement h]
\centering
  \includegraphics[height=2.in,width=2.75in,angle=0]{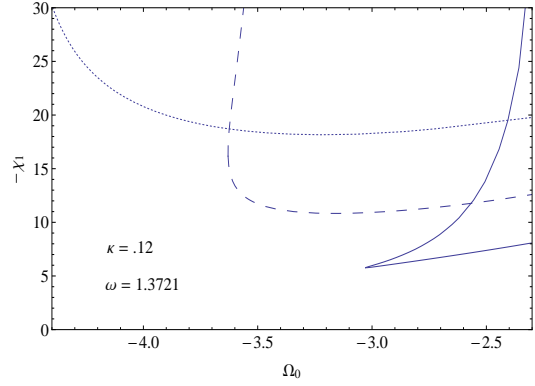}
\caption {$-\chi_1(\tilde\Omega_0)$ [with  $\kappa=.12$ and  $\omega=1.3721$] is plotted for the solutions having monotonically decreasing $\chi$  (solid curve), the single node solutions (dashed curve) and the double node solutions (dotted curve).   $\chi_1(\tilde\Omega_0)$ is seen to have a  cusp singularity at $\tilde\Omega_0\approx -3.02$ for the zero node solutions. }  
\label{fig:test}
\end{figure}
\begin{figure}[placement h]
\centering
\begin{subfigure}{.5\textwidth}
  \centering
  \includegraphics[height=2.in,width=2.5in,angle=0]{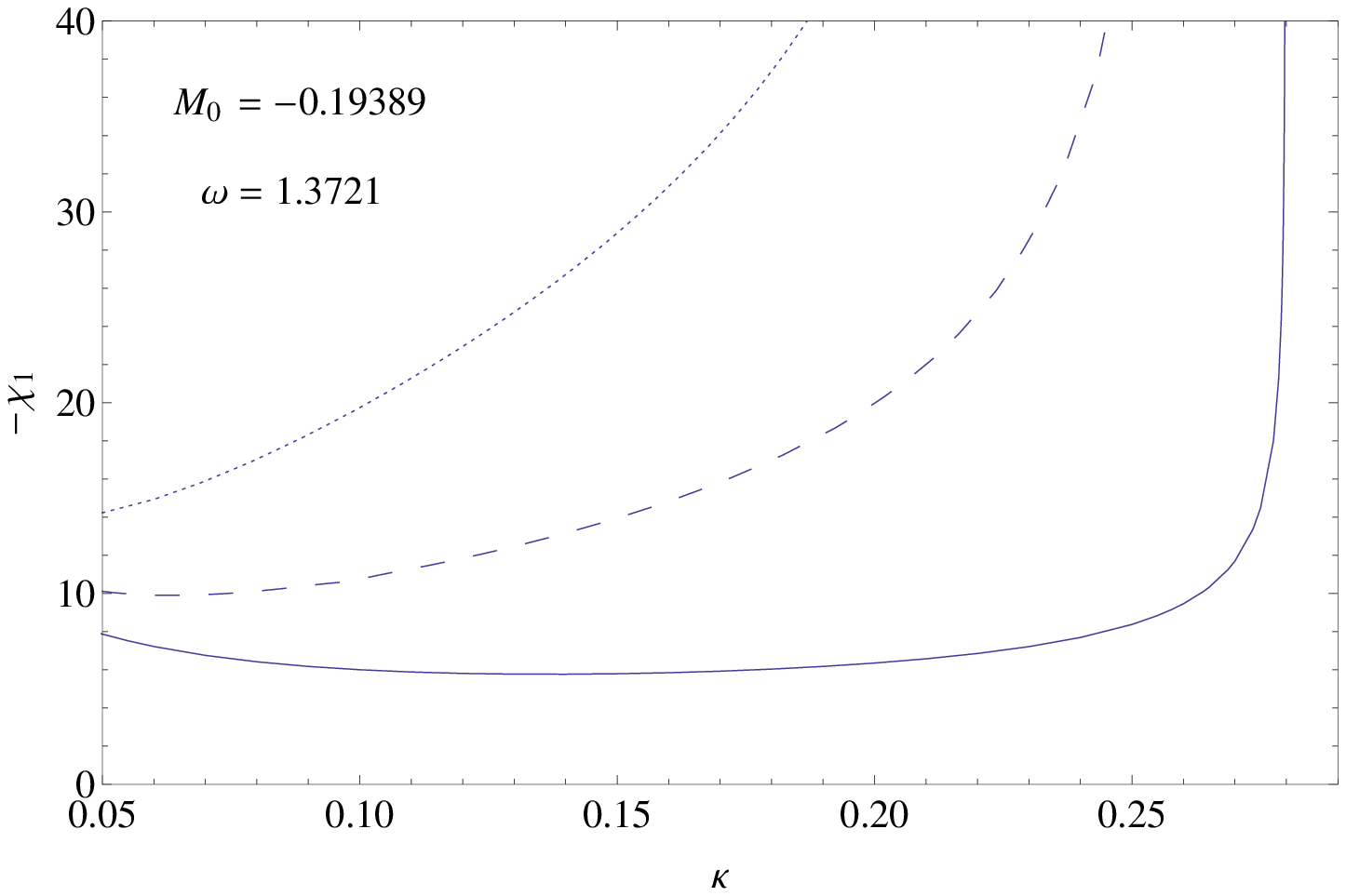}
  \caption{}
  \label{fig:sub1}
\end{subfigure}%
\begin{subfigure}{.5\textwidth}
  \centering
  \includegraphics[height=2.in,width=2.5in,angle=0]{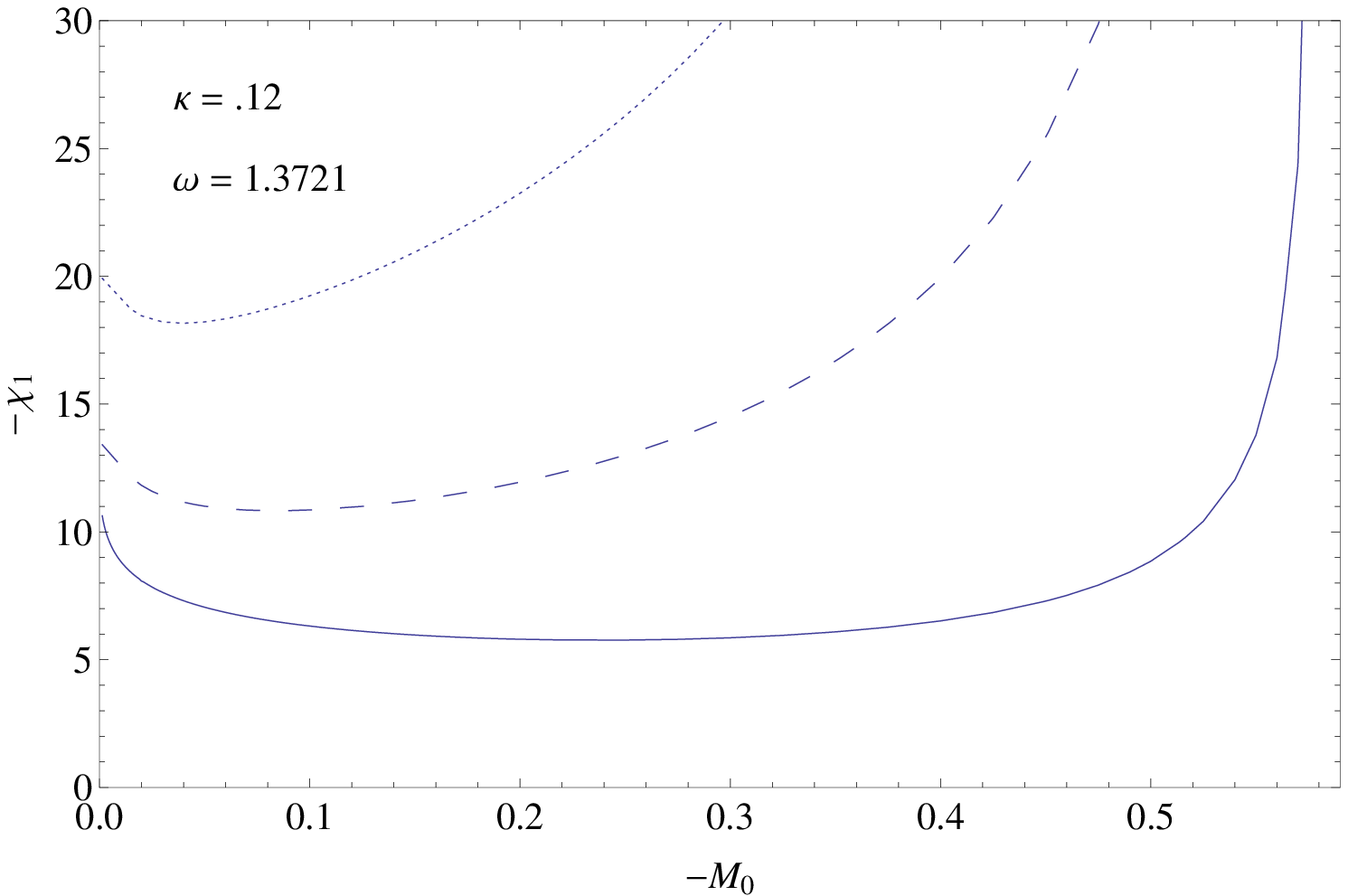}
  \caption{}
  \label{fig:sub1}
\end{subfigure}%
\caption {In (a) ,$-\chi_1(\kappa)$ [with $\omega\approx 1.3721$ and $M_0\approx -.019389$] is plotted  for the solutions having monotonically decreasing $\chi$  (solid curve), the single node solutions (dashed curve) and the double node solutions (dotted curve). The maximum value for $\kappa$ for the zero node solutions is approximately $.28$. In (b), $-\chi_1(-M_0)$ [with $\omega\approx 1.3721$ and $\kappa\approx .12$] is plotted for  the solutions having monotonically decreasing $\chi$  (solid curve), the single node solutions (dashed curve) and the double node solutions (dotted curve). The maximum value for $-M_0$ for the zero node solutions is approximately $.58$.}  
\label{fig:test}
\end{figure}

\section{Collective coordinate quantization}

The collective coordinate quantization of the soliton allows for an alternative definition of the mass and angular momentum of the soliton.  We denote them by $ {\cal M}$ and $ {\cal J}$, respectively.  Both can be computed from the action (\ref{srtngactn}) evaluated for the soliton.  A Chern-Simons term\cite{Bowick:1985ua},\cite{Balachandran:1991zj} can  be included in  the total action, and this will produce a  contribution which is linear in the angular velocity, in addition to those coming from (\ref{srtngactn}).  However, such contributions do not affect the energy spectrum, and so we will not consider the Chern-Simons term.

In the collective coordinate approach one replaces $\tilde\omega$ with a dynamical angular velocity $\dot\psi$, with the caveat that its variation is sufficiently small so that it doesn't significantly change the values of the mass $ {\cal M}$ or the moment of inertia ${\cal I}$ of the soliton. $ {\cal M}$ is defined as the  $\dot\psi$-independent contribution to the soliton action, while ${\cal I}/2$  is the coefficient of the quadratic contribution in $\dot\psi$. As indicated above, there is also a linear contribution.  Thus the soliton action can be written
\be S=\int dt\;\Bigl\{\frac 12{\cal  I}\dot \psi^2+ \alpha\dot \psi- {\cal M} \Bigr\} \;, \ee
where ${\cal I}$, $\alpha$ and ${\cal M}$ are given by the radial integrals
\beqa
{\cal I}&=&  2\pi \int dx\, \frac{ x \sqrt{B}}A\sin^2\chi\cr &&\cr
\alpha &=&  2\pi \int dx\, \frac{ x\tilde\Omega \sqrt{B}}A\sin^2\chi\cr &&\cr
{\cal M}&=&-\frac {\pi}{\kappa}\int\frac{ dx}{\sqrt{B}} \;\biggl\{ { A'}+\frac { x^3\tilde\Omega'^2}{2}+{2 xB} - {  x\kappa}\biggl( A\chi'^2+\Bigl(\frac 1{x^2}-\frac{ \tilde\Omega^2}{A}\Bigr){B}\sin^2\chi\biggr)\;-4 x\sqrt{B} \Biggr\} \;.\eeqa
The infinite $AdS$ vacuum action $S_{AdS}$ was subtracted from ${\cal M}$. The angular momentum  of the soliton is ${\cal J}= {\cal I}\dot\psi$.  
From the asymptotic behavior (\ref{asmptcfrm}) as $x\rightarrow\infty$  and the behavior (\ref{nrrignfrm}) or (\ref{fntnrrign}) as $x\rightarrow 0$,  the integral expressions for ${\cal I}$, $\alpha$ and ${\cal M}$ are finite. (This is in contrast to  the moment of inertia for the $\sigma$-model soliton in  Minkowski space-time, which is not bounded, leading to a  spontaneous breakdown of rotational symmetry.\cite{Stern:1987di})  For the  $n=1$ solution with $M=J=1$  appearing in figures 1 through 5
we get  ${\cal M}\approx -22.1$ and ${\cal J}={\cal I}\tilde\omega\approx 7.6$.  For the topological soliton  illustrated in figure 6 having $M\approx .0539$ and $J\approx .0817$ we get  ${\cal M}\approx 2.1$ and ${\cal J}={\cal I}\tilde\omega\approx 2.78 $.
 
The  Hamiltonian for the system is 
\be H=\frac{{\cal J}^2}{2{\cal I}}  +{\cal M}\;.\label{hamcc}\ee
The angular momentum ${\cal J}$ is related to the canonical momentum $p_\psi$ by $ {\cal J}=p_\psi -\alpha$.
   Its  Poisson bracket with the $U(1)$ phase $e^{i\psi}$ is then
\be\{e^{i\psi}, {\cal J}\}=i e^{i\psi}\;.\ee

In passing to the quantum theory the spectrum of the  operator  $\hat {\cal J}$ corresponding to $ {\cal J}$ is not unique, the eigenvalues being  integers plus an arbitrary constant.\cite{Bowick:1985ua},\cite{Balachandran:1991zj}
 It obeys the commutator
\be [\widehat{e^{i\psi}},\hat {\cal J}]=-\hbar\, \widehat{e^{i\psi}}\;,\ee
where $\widehat{e^{i\psi}}$ is the  operator corresponding to ${e^{i\psi}}$.  The algebra has the  Casimir operator
$\;\exp{\frac {2\pi i}\hbar \hat {\cal J}}$, whose eigenvalues are phases $e^{i\phi_0}$ which label different irreducible representations in the quantum theory.  The spectrum for $\hat {\cal J}$ is then $\hbar$ times an integer $m$ plus an arbitrary phase constant, $ \hbar m +\frac{\phi_0\hbar}{2\pi} $, and so from (\ref{hamcc}) the energy eigenvalues are \be
E_m=\frac{\hbar^2}{2{\cal I}}\Bigl(  m +\frac{\phi_0}{2\pi}\Bigr)^2  +{\cal M}\ee 
Of course the energy spectrum depends on an additional integer $n$, the winding number, since ${\cal I}$ and ${\cal M}$ do.

\section{The question of hairy BTZ black hole solutions}

The functions $A(x)$ and $B(x)$ were positive for all of the numerical solutions  obtained previously by integrating either from $x\rightarrow 0$ or $x\rightarrow\infty$.  Thus none of these solutions developed horizons.
We can  instead assume a priori the existence of at least one horizon. In the case of multiple horizons, let $x_H>0$ denote the the location outer most one.    Then $A(x_H)=0$. 
A consistent solution of  (\ref{eqsdmrnqdratic})  near the horizon, $x-x_H<<1$, can be obtained by demanding that $\tilde\Omega(x_H)=-\tilde\omega $.  A power series expansion for the functions $A$, $B$, $\tilde\Omega$ and $\chi$ can then be determined from three independent parameters, say  $B_H=B(x_H)$, $\chi_H=\chi(x_H)$ and $\tilde\Omega_1=\tilde\Omega'(x_H)$, as well as $x_H$.  Up to first order in  $x-x_H$,
 \beqa A &\rightarrow & A_1\,(x-x_H)+{\cal O}\Bigl((x-x_H)^2\Bigr) \cr &&\cr  B &\rightarrow & B_H+B_1\,(x-x_H)+{\cal O}\Bigl((x-x_H)^2\Bigr)  \cr &&\cr \tilde\Omega  &\rightarrow & -\tilde\omega  +\tilde\Omega_1\,(x-x_H)+{\cal O}\Bigl((x-x_H)^2\Bigr)\cr &&\cr 
\chi  &\rightarrow &\chi_H+\chi_1\,(x-x_H)+{\cal O}\Bigl((x-x_H)^2\Bigr)\;,\label{bhvrnrhrzn}\; \eeqa where the horizon parameters $ A_1$, $B_H$, $B_1$, $\tilde\Omega_1$,  $\chi_H$ and $\chi_1$  satisfy
\beqa A_1&=&2 B_H x_H -\frac{\tilde\Omega_1^2 x_H^3}2-\frac{\kappa B_H\sin^2\chi_H}{x_H}\cr &&\cr B_1&=& {2\kappa x_HB_H }\Bigl( \chi_1^2 +\frac {B_H \tilde\Omega_1^2 \sin^2\chi_H}{A_1^2}\Bigr)\cr &&\cr \chi_1&=&\frac{B_H \sin(2\chi_H)}{2 x_H^2A_1}\;.\label{bndvlsatxH}\eeqa
Setting $\chi$ at $x=x_H$ equal to an integer multiple of $\pi$ in (\ref{bndvlsatxH}) leads to $\chi_1=0$, along with the vanishing of higher derivatives of $\chi$ at  $x=x_H$.
Therefore there are no solutions for this case, and so the domain $x\ge x_H$ of the nonlinear $\sigma$-model cannot be taken to be $S^2$.  Since  we desire no horizons  in the  domain $x\ge x_H$, we require that
$A(x), B(x)>0$  in this  domain and so  $A_1>0$.  Therefore,  $ B_H(1-\frac{\kappa \sin^2\chi_H}{2x_H^2}) > \frac{\tilde\Omega_1^2 x_H^2}4$ and  $ { \sin^2\chi_H}<\frac{2x_H^2} \kappa$.

The above conditions are of course satisfied for the `bald' BTZ solution where the four functions
 $A$, $B$, $\tilde\Omega$ and $\chi$ are respectively
\be A_{BTZ}=x^2-M+ \frac{J^2}{x^2}\quad \qquad B_{BTZ}=1\qquad\quad \tilde\Omega_{BTZ}=\tilde\Omega_\infty+ \frac{J}{x^2}\quad\qquad\chi_{BTZ}=0\;.\ee 
Then identifying $x_H$ with the outer horizon, $x_H^2= \frac 12 \left(M+\sqrt{M^2-4J^2}\right)$, we get the following results for the horizon parameters
\beqa  &&A_1= \frac 2{x_H}\sqrt{M^2-4J^2}\qquad\quad B_H=1 \qquad\quad\tilde\Omega_1=- \frac{2J}{x_H^3}\cr &&\cr
&&\tilde\Omega_\infty+\tilde\omega+\frac{J}{x_H^2}\;=\;B_1\;=\;\chi_H\;=\;\chi_1=0\;.\label{twenty2}
\eeqa

For a hairy black hole we drop the restriction of  $\chi_H=\chi_1=0$ as well as the other conditions in (\ref{twenty2}) that yield the BTZ solution. A monotonically decreasing function $\chi(x)$ and monotonically increasing function for $A(x)$ require $\chi_1<0$ and $A_1>0$, respectively. From the last equation in (\ref{bndvlsatxH}) and $B_H>0$ it follows that $\sin (2\chi_H)<0$.   Given these inequalities on the horizon parameters, along with the conditions (\ref{bndvlsatxH}), we can then integrate the equations of motion  (\ref{eqsdmrnqdratic}) from $x_H$ to $ x\rightarrow\infty$.  Upon so doing we were unable to recover the asymptotic solution (\ref{asmptcfrm})  at $ x\rightarrow\infty$, and hence we did not find any black hole solutions with nonlinear $\sigma$-model hair.

\section{Concluding remarks}

We obtained numerical solutions for two types, $i)$ and $ii)$, of rotating self-gravitating topological  solitons of the  nonlinear $\sigma$-model where the space-time approaches $AdS^3$ in the large distance limit.  Upon embedding the solutions in $3+1$ dimensions, they can be interpreted as cosmic strings.  For  the type $i)$ solution, any time slice of the space-time domain has a causal singularity, which is analogous to the BTZ black hole singularity.  On the other hand, the space-time domain is singularity free for  type $ii)$ solutions. $\chi(x)$ for such solutions exhibit an arbitrary number of nodes.  No evidence of a horizon was seen for any of the solutions.   Therefore these solutions are not hairy black holes, and furthermore  the type $i)$ solutions have naked singularities. 

Among the lines of inquiry that  remain to be investigated is the search for black hole solutions with nonlinear $\sigma$-model hair, analogous to the known $3+1$ dimensional black hole solutions with Skyrme hair.  This may require the inclusion of higher order derivative terms, analogous to the Skyrme term,  in the nonlinear $\sigma$-model action.   While the solitons obtained here are topologically stable, the question of whether or not they  are stable under local fluctuations needs to be determined.  Finally, it is worthwhile to understand  the role that these new three-dimensional $AdS$ solutions may or may not play for the two-dimensional space-time  boundary
field theory.


\bigskip

\begin{thebibliography}{99}
\bibitem{Witten:1998zw} 
  E.~Witten,
  ``Anti-de Sitter space, thermal phase transition, and confinement in gauge theories,''
  Adv.\ Theor.\ Math.\ Phys.\  {\bf 2}, 505 (1998).



\bibitem{Banados:1992gq}  M.~Banados, C.~Teitelboim and J.~Zanelli,
 ``The Black hole in three-dimensional space-time,''
  Phys.\ Rev.\ Lett.\  {\bf 69}, 1849 (1992)
  doi:10.1103/PhysRevLett.69.1849;
  M.~Banados, M.~Henneaux, C.~Teitelboim and J.~Zanelli,
  ``Geometry of the (2+1) black hole,''
  Phys.\ Rev.\ D {\bf 48}, 1506 (1993)
  Erratum: [Phys.\ Rev.\ D {\bf 88}, 069902 (2013)].


\bibitem{Horowitz:1998ha} 
  G.~T.~Horowitz and R.~C.~Myers,
 ``The AdS / CFT correspondence and a new positive energy conjecture for general relativity,''
  Phys.\ Rev.\ D {\bf 59}, 026005 (1998).

\bibitem{Henneaux:2002wm} 
  M.~Henneaux, C.~Martinez, R.~Troncoso and J.~Zanelli,
  ``Black holes and asymptotics of 2+1 gravity coupled to a scalar field,''
  Phys.\ Rev.\ D {\bf 65}, 104007 (2002).

\bibitem{Banados:2005hm} 
  M.~Banados and S.~Theisen,
  ``Scale invariant hairy black holes,''
  Phys.\ Rev.\ D {\bf 72}, 064019 (2005).


\bibitem{Brihaye:2013tra} 
  Y.~Brihaye, B.~Hartmann and S.~Tojiev,
  ``AdS solitons with conformal scalar hair,''
  Phys.\ Rev.\ D {\bf 88}, 104006 (2013).

\bibitem{Anabalon:2016izw} 
  A.~Anabalon, D.~Astefanesei and D.~Choque,
  ``Hairy AdS Solitons,''
  arXiv:1606.07870 [hep-th].


\bibitem{Bizon:2004wa} 
  P.~Bizon and A.~Wasserman,
``A Note on the non-existence of sigma-model solitons in the 2+1 dimensional AdS gravity,''
  Phys.\ Rev.\ D {\bf 71}, 108701 (2005).

\bibitem{Clement:1976hh} 
  G.~Clement,
  ``Field-Theoretic Extended Particles in Two Space Dimensions,''
  Nucl.\ Phys.\ B {\bf 114}, 437 (1976).

\bibitem{Heusler:1991xx} 
  M.~Heusler, S.~Droz and N.~Straumann,
 ``New black hole solutions with hair,''
  Phys.\ Lett.\ B {\bf 268}, 371 (1991);
  ``Stability analysis of selfgravitating skyrmions,''
  Phys.\ Lett.\ B {\bf 271}, 61 (1991);
 M.~Heusler, N.~Straumann and Z.~h.~Zhou,
 ``Selfgravitating solutions of the Skyrme model and their stability,''
  Helv.\ Phys.\ Acta {\bf 66}, 614 (1993).

\bibitem{Ioannidou:2006nn} 
  T.~Ioannidou, B.~Kleihaus and J.~Kunz,
  ``Spinning gravitating skyrmions,''
  Phys.\ Lett.\ B {\bf 643}, 213 (2006).

\bibitem{Glendenning:1988qy} 
  N.~K.~Glendenning, T.~Kodama and F.~R.~Klinkhamer,
  ``Skyrme Topological Soliton Coupled To Gravity,''
  Phys.\ Rev.\ D {\bf 38}, 3226 (1988).


\bibitem{Piette:2007wd} 
  B.~M.~A.~G.~Piette and G.~I.~Probert,
  ``Towards skyrmion stars: Large baryon configurations in the Einstein-Skyrme model,''
  Phys.\ Rev.\ D {\bf 75}, 125023 (2007).

\bibitem{Nelmes:2011zz} 
  S.~Nelmes and B.~M.~A.~G.~Piette,
  ``Skyrmion stars and the multilayered rational map ansatz,''
  Phys.\ Rev.\ D {\bf 84}, 085017 (2011).

\bibitem{Luckock:1986tr} 
  H.~Luckock and I.~Moss,
  ``Black Holes Have Skyrmion Hair,''
  Phys.\ Lett.\ B {\bf 176}, 341 (1986).

\bibitem{Bizon:1992gb} 
  P.~Bizon and T.~Chmaj,
 ``Gravitating skyrmions,''
  Phys.\ Lett.\ B {\bf 297}, 55 (1992);
  ``Critical collapse of Skyrmions,''
  Phys.\ Rev.\ D {\bf 58}, 041501 (1998).


\bibitem{Kleihaus:1995vq} 
  B.~Kleihaus, J.~Kunz and A.~Sood,
  ``SU(3) Einstein-Skyrme solitons and black holes,''
  Phys.\ Lett.\ B {\bf 352}, 247 (1995).

\bibitem{Tamaki:2001wca} 
  T.~Tamaki, K.~i.~Maeda and T.~Torii,
  ``Internal structure of Skyrme black hole,''
  Phys.\ Rev.\ D {\bf 64}, 084019 (2001).


\bibitem{Sawado:2004yq} 
  N.~Sawado, N.~Shiiki, K.~i.~Maeda and T.~Torii,
  ``Regular and black hole Skyrmions with axisymmetry,''
  Gen.\ Rel.\ Grav.\  {\bf 36}, 1361 (2004);
  N.~Shiiki and N.~Sawado,
  ``Regular and black hole solutions in the Einstein-Skyrme theory with negative cosmological constant,''
  Class.\ Quant.\ Grav.\  {\bf 22}, 3561 (2005);``Black hole skyrmions with negative cosmological constant,''
  Phys.\ Rev.\ D {\bf 71}, 104031 (2005);  ``Black holes with skyrme hair,''
  gr-qc/0501025.


\bibitem{Brihaye:2005an} 
  Y.~Brihaye and T.~Delsate,
  ``Skyrmion and Skyrme-black holes in de Sitter spacetime,''
  Mod.\ Phys.\ Lett.\ A {\bf 21}, 2043 (2006).

\bibitem{Nielsen:2006gb} 
  A.~B.~Nielsen,
 ``Skyrme Black Holes in the Isolated Horizons Formalism,''
  Phys.\ Rev.\ D {\bf 74}, 044038 (2006).

\bibitem{Duan:2007df} 
  Y.~S.~Duan, X.~H.~Zhang and L.~Zhao,
 ``Topological aspect of black hole with Skyrme hair,''
  Int.\ J.\ Mod.\ Phys.\ A {\bf 21}, 5895 (2006).

\bibitem{Doneva:2011gx} 
  D.~D.~Doneva, I.~Z.~Stefanov and S.~S.~Yazadjiev,
  ``Solitons and Black Holes in a Generalized Skyrme Model with Dilaton-Quarkonium field,''
  Phys.\ Rev.\ D {\bf 83}, 124007 (2011);   D.~D.~Doneva, K.~D.~Kokkotas, I.~Z.~Stefanov and S.~S.~Yazadjiev,
 ``Time Evolution of the Radial Perturbations and Linear Stability of Solitons and Black Holes in a Generalized Skyrme Model,''
  Phys.\ Rev.\ D {\bf 84}, 084021 (2011).

\bibitem{Gibbons:2010cr} 
  G.~W.~Gibbons, C.~M.~Warnick and W.~W.~Wong,
  ``Non-existence of Skyrmion-Skyrmion and Skrymion-anti-Skyrmion static equilibria,''
  J.\ Math.\ Phys.\  {\bf 52}, 012905 (2011).

\bibitem{Canfora:2013osa} 
  F.~Canfora and H.~Maeda,
  ``Hedgehog ansatz and its generalization for self-gravitating Skyrmions,''
  Phys.\ Rev.\ D {\bf 87},  084049 (2013).

\bibitem{Dvali:2016mur} 
  G.~Dvali and A.~Gussmann,
 ``Skyrmion Black Hole Hair: Conservation of Baryon Number by Black Holes and Observable Manifestations,''
  arXiv:1605.00543 [hep-th].

\bibitem{Adam:2016vzf} 
  C.~Adam, O.~Kichakova, Y.~Shnir and A.~Wereszczynski,
  ``Hairy black holes in the general Skyrme model,'' Phys.\ Rev.\ D {\bf 94}, 024060 (2016).

\bibitem{Gudnason:2016kuu} 
  S.~B.~Gudnason, M.~Nitta and N.~Sawado,
  ``Black hole Skyrmion in a generalized Skyrme model,''
  arXiv:1605.07954 [hep-th]. 

\bibitem{York:1972sj} 
  J.~W.~York, Jr.,
  ``Role of conformal three geometry in the dynamics of gravitation,''
  Phys.\ Rev.\ Lett.\  {\bf 28}, 1082 (1972); G.~W.~Gibbons and S.~W.~Hawking,
  ``Action Integrals and Partition Functions in Quantum Gravity,''
  Phys.\ Rev.\ D {\bf 15} (1977) 2752.


\bibitem{Bowick:1985ua}
  M.~J.~Bowick, D.~Karabali and L.~C.~R.~Wijewardhana,
  ``Fractional Spin via Canonical Quantization of the O(3) Nonlinear Sigma Model,''
  Nucl.\ Phys.\ B {\bf 271} (1986) 417.


\bibitem{Balachandran:1991zj} 
  A.~P.~Balachandran, G.~Marmo, B.~S.~Skagerstam and A.~Stern,
  ``Classical topology and quantum states,''
  Singapore, Singapore: World Scientific (1991) 358 p.

\bibitem{Stern:1987di} 
  A.~Stern,
  ``Frozen Solitons in a Two-dimensional Ferromagnet,''
  Phys.\ Rev.\ Lett.\  {\bf 59}, 1506 (1987).


\end{thebibliography}
 \end{document}